\documentclass[nofootinbib,aps,11pt,showpacs]{revtex4-1}
\usepackage{graphics}
\usepackage{graphicx}
\usepackage{cancel}
\usepackage{amssymb}
\usepackage{textcomp}
\usepackage{amsmath}
\usepackage{bm}
\usepackage{times}
\usepackage{epsfig}
\usepackage{color}
\usepackage{slashed}
\usepackage{multirow}
\usepackage{axodraw}

\newcommand{\rd}{\frac{1}{\sqrt{2}}}
\newcommand{\tw}{\theta_W}
\newcommand{\wbp}{W^+}
\newcommand{\wbm}{W^-}
\newcommand{\nub}{\overline{\nu}}
\newcommand{\lp}{\ell^+}
\newcommand{\lm}{\ell^-}

\begin{document}

\begin{flushright}
MADPH-11-1571
\end{flushright}

\title{\LARGE Phenomenology of A Lepton Triplet }
\bigskip
\author{Antonio Delgado$^1$, Camilo Garcia Cely$^2$, Tao Han$^2$, Zhihui Wang$^{2,3}$ }
\affiliation{
$^1$Department of Physics, University of Notre Dame, Notre Dame, IN 46556, USA\\
$^2$Department of Physics, University of Wisconsin, Madison, WI 53706, USA \\
$^3$Department of Physics, Harbin Institute of Technology, Harbin 150001, P.R.~China } 
\date{\today}

\begin{abstract}

The most general phenomenological model involving a lepton triplet with hypercharge $\pm 1$ is constructed. A distinctive feature of this model is the prediction of a doubly charged lepton, and a new heavy Dirac neutrino. We study the phenomenology of these exotic leptons in both low-energy experiments and at the LHC. The model predicts FCNC processes such as muon and tau rare decays, which are studied in detail in order to constrain the model parameters. All the decay channels of the exotic leptons are described for a wide range of parameters. 
It is found that, 
if the mixing parameters between the exotic and light leptons are not too small ($>10^{-6}$), then
they can be observable to a $3-5\sigma$ statistical significance at the 7 TeV LHC with $10-50$ fb$^{-1}$ luminosity
for a 400 GeV mass, and 14 TeV with $100-300$ fb$^{-1}$ luminosity for a 800 GeV mass.

\end{abstract}
\pacs{14.60.Pq, 14.60.Hi, 14.60.St}
\maketitle

\section{Introduction}

Although the success of the Standard Model (SM) as a way to correctly describe the interactions among particles is beyond any doubt, there are good reasons to believe that the SM is not the ultimate theory to describe Nature. Particle dark matter, neutrino masses, and the actual mechanism of electroweak breaking are among the pressing issues. Especially the nature of the sector which breaks the EW symmetry and gives masses to gauge bosons and fermions has been one of the leading motivations for many theoretical considerations beyond the standard model. Supersymmetry tries to explain the hierarchy problem while having a weakly coupled Higgs sector, warped extra dimensions~\cite{Randall:1999vf} on the other hand provide with an alternative explanation which, through the AdS/CFT correspondence~\cite{adscft}, can be understood as being dual to a strongly coupled origin for the Higgs~\cite{rscft}. In particular one can accommodate the old idea of having the Higgs as a pseudo-Goldstone boson of a global symmetry~\cite{Weinberg:1972fn} in this framework by having a model where the Higgs comes from a gauge multiplet, the so-called gauge-Higgs unification~\cite{Contino:2003ve}.

Some model building is required in order to construct a complete model of gauge-Higgs unification that passes all the electroweak precision tests~\cite{tests}, like gauging $SU(2)_R$ in order to protect the $\rho$-parameter or including special representations for fermions to cancel dangerous contributions to $Z\to b\overline{b}$. One of the consequences of that is the appearance of extra fermionic states with exotic hypercharges. This paper deals with the phenomenology of some of those exotic states following previous studies~\cite{Contino:2008hi}. Specifically we will study the phenomenology of a vector-like triplet of leptons with $Y=1$ that mixes with the usual leptons of the SM via a Yukawa coupling with the Higgs. 

Perhaps the most interesting consequence of this model is the existence of a doubly charged lepton, which we will
refer to an exotic heavy lepton.
Similar new leptonic states have been considered ~\cite{Chua:2010me}, most notably in the context of a possible mechanism to generate neutrino masses, the so-called Type III see-saw~\cite{Foot:1988aq}, although in that particular model the lepton introduced had $Y=0$. Doubly charged fermions have been also studied as doubly charged Higgsinos \cite{higgsinos} in the context of an extended SUSY theory or flavor models in warped dimensions \cite{jose} or in more general models \cite{Strumia}.
Here we will follow a model-independent approach. 
We will introduce the most general Lagrangian including this triplet. The mass will be treated as a free parameter. We will introduce general mixing matrices among these new states and the SM particles. Upon diagonalization of the mass matrices 
the couplings of these extra leptons to the SM particles will be bounded by experiments on FCNC and neutrino physics. We will find that the absence of exotic decays of the muon will put the stringiest bounds on those couplings.
Under those constraints, we will then study the decay widths and channels for these new particles. We will then perform an analysis of the possible signatures and SM backgrounds for discovery of these exotic states at the LHC.

The paper is organized as follows, in section II we present the model with particular emphasis to the spectrum and interactions of these new fields. Section III is devoted to the constraints that different experimental facts put in the model. We study the decay patterns on Section IV that leads to the different signatures in Section V. Our conclusions are presented in Section VI whereas we have relegated some technical details to the Appendix.

\section{Description of the model}

\subsection{The model for one generation}

In this model, there is a vector-like $SU(2)$ triplet of exotic leptons with hypercharge $Y=\pm 1$. 
A singlet right handed neutrino is also included to give mass to the neutrino. As a result the particle content, according to the
$(SU(2),U(1)_Y)$ quantum numbers, is
\begin{eqnarray}
	H &=& \begin{pmatrix} \phi^+\\ \phi^0 \end{pmatrix} \in (2,1/2), \quad e_R \in (1,-1),\quad \nu_R \in (1,0),\quad
	L_L = \begin{pmatrix} \nu\\ e \end{pmatrix}_L \in (2,-1/2)  \\ \nonumber
	X_L &=& \begin{pmatrix} X^{0}\\X^{-}\\X^{--} \end{pmatrix}_L , \quad X_R = \begin{pmatrix} X^{0}\\X^{-}\\X^{--} \end{pmatrix}_R \in (3, -1),
\end{eqnarray}
where $L,R$ refer to the chirality of the fermions. 

The most general lagrangian that gives rise to the lepton masses without breaking gauge invariance or lepton number is
\begin{eqnarray}
 -{\cal L}_Y &=& \lambda_1 \overline{L_L} H {e_R} + \lambda_2 \overline{L_L} H^c {\nu _R} + \lambda_3 \overline{X_R} H^c L_L + M_1 \overline{X}X + h.c. 
\end{eqnarray}
We are focusing on Dirac Leptons and therefore majorana mass terms are not allowed. For sake of simplicity, we work in the unitary gauge here and leave the general case for Appendix C.  We thus have $H=\begin{pmatrix} 0 \\ \frac{v+h(x)}{\sqrt2}\end{pmatrix}$ and
\begin{eqnarray}
	\nonumber
	- {\cal L}_Y 
	   &=& \lambda_1 \frac{ v + h(x) }{\sqrt2}  \overline{e_L} e_R + \lambda_2 \frac{ v +h(x) }{\sqrt2}  \overline{\nu_L} \nu_R 
	  +  \lambda_3 \frac{v+h(x)}{2}  \overline{X^-_R} e_L + \lambda_3 \frac{v+h(x)}{2} \overline{X^0_R}  \nu_L \\ 
	   && + M_1 ( \overline{X^{--_L}} X^{--}_R +\overline{X^-_L} X^-_R+\overline{X^0_L} X^0_R)+ h.c. 
\end{eqnarray}
Setting $m_i = \rd v \lambda_i$,  we can rewrite the lagrangian as
\begin{eqnarray} 
	-{\cal L} _Y 
	   &=& \overline{N_L} M_N N_R + \overline{E_L} M_E E_R + \frac{1}{v} h(x) \overline{E_L} M_E\vert_{M_1=0} E_R  \nonumber\\
	    && +  \frac{1}{v} h(x) \overline{N_L} M_N \vert_{M_1=0} N_R +  M_1 \overline{X^{--}_L} X^{--}_R +h.c.,
	   \label{Ly}
\end{eqnarray}
where the neutral and charged leptons as well as their mass matrices are 
\begin{eqnarray}
	N = \begin{pmatrix} \nu \\ X^0  \end{pmatrix}, &\hfill& \;\;\;\;\;
	E = \begin{pmatrix} e \\ X^-   \end{pmatrix} \label{Col} \\
	M_N =\begin{pmatrix} m_2  &\hfill& m_3*\\
	                             0    &\hfill&  M_1 \end{pmatrix}, &\hfill&\;\;\;\;\;
	M_E =\begin{pmatrix} m_1  &\hfill& \frac{m_3*}{\sqrt{2}} \\
	                             0    &\hfill&  M_1 \end{pmatrix}, 
	\label{M}
\end{eqnarray}
and $M_N \vert_{M_1=0}$ means setting $M_1 = 0$ in the matrix defined above for $M_N$. Similar remarks apply for $M_E$. 

\subsection{Generalization to three generations}

It is straightforward to generalize equations (\ref{Ly}), (\ref{Col}) and (\ref{M}) to include three generations of $SU(2)$ doublets. The mass matrices can then be diagonalized by biunitary transformations
\begin{eqnarray}
	S_E^\dagger M_E T_E &=& M_{Ed} = diag(m_e, m_\mu, m_\tau, M_2), \ \ 
	S_N^\dagger M_N T_N = M_{Nd} = diag(m_{\nu_1}, m_{\nu_2}, m_{\nu_3}, M_3),~~~
	\label{SDT}
\end{eqnarray}
where, as usual, the matrices $S_E$, $T_E$, $S_N$ and $T_N$ are unitary and the diagonal elements are the tree level masses. On dimensional grounds we expect $ M_2 - M_1 \varpropto m_{e_i}^2/M_1$ and $ M_3 - M_1 \varpropto m_{\nu_i}^2/M_1$. Hence, the masses of the leptons are nearly degenerate at tree-level. We will see later that quantum corrections lift this degeneracy. 

Similar to a general fermionic sector with arbitrary Yukawa couplings, there are more theory parameters (9 elements of $\lambda_1$, 9 elements of $\lambda_2$, 3 elements of $\lambda_3$ and $M_1$) than those that can be experimentally determined (masses of the leptons and their mixings such as the PMNS matrix elements). Thus it is necessary to parameterize the fermionic sector by a few more physical parameters. We find convenient to introduce
\begin{eqnarray}
	V = S_N^\dagger S_E \hspace{20pt}
	v_E &=& S_E^\dagger \begin{pmatrix} 0 \\1 \end{pmatrix} \hspace{20pt} 	
	v_N = V v_E \label{vdefap}.
\end{eqnarray}
The $4\times 1$ matrices $v_{E}, v_{N}$ characterize the mixing among the SM leptons and the new heavy triplet
in the gauge interactions. 
In appendix A, we derive the following relations:
\begin{eqnarray}
	\vert v_{E4} \vert =M_1/M_2 \hspace{20pt} &\;\;\;\;\;\;\;\;\;&
	\vert v_{N4} \vert =M_1/M_3 \nonumber\\
	V_{i4} = \left( \rd \frac{M_3^2}{M_1^2} \delta_{i4} + 1 - \rd \right) v^*_{E4} v_{Ni} &\;\;\;\;\;\;\;\;\;&
	V_{4i}^* = \sqrt{2} \left(\frac{M_2^2}{M_1^2} \delta_{i4} - 1+\rd \right) v_{N4}^* v_{Ei}.
	\label{newcon}
\end{eqnarray}
Several remarks are in order:

$\bullet$ First, according to relations (\ref{newcon}), our physical parameters are the masses of the leptons $m_e, m_\mu, m_\tau,$ $m_{\nu_1}, m_{\nu_2}, m_{\nu_3}, M_1, M_2$ and $M_3$  plus the mixings $v_{Ei}$ and $V_{ij}$ with $i,j =1, 3$ (which will be related to the PMNS matrix in section II.D). Because of the way we constructed them, they are independent of each other. As a result, there are no relations among the neutrino masses in our model. This is consistent with the fact that all of our neutrinos are of Dirac type.

$\bullet$ Second, because the near degeneracy of $M_{1},M_{2},M_{3}$, equations (\ref{newcon}) imply $v_{E}^T \approx v_{N}^T \approx (0,0,0,1)$.
Consequently, the mixing elements $V_{4i}^*$ and $V_{i4}$ for $i=1,2,3$ are very small.
Thus, the matrix $V$ decomposes into two blocks: a 3x3 unitary matrix and a number $1$. This limit corresponds to the situation in which the Standard Model leptons do not interact with the exotic triplet directly. 

$\bullet$ Third,  these relations imply that $M_1$ should be smaller than $M_2$ and $M_3$. In other words, the doubly charged lepton should be lighter than the singly charged lepton and the neutral lepton at \textit{tree} level. This situation changes once quantum corrections are taken into account, as will be discussed in the next section. 

$\bullet$ The peculiar factor $1-\frac{1}{\sqrt2}$ on equations (\ref{newcon}), which comes from equation (\ref{trivsdou}), can be traced back to the particular form of the mass matrices (\ref{M}). This in turn can be associated to the triplet nature of the exotic leptons. If they constituted a doublet representation instead, there would not be a $1/\sqrt{2}$ on (\ref{M}) and the factor $1-\frac{1}{\sqrt{2}}$ would be absent. As we will see later in section II.D (or in appendix B),  this fact will be crucial to show that coupling among the neutral exotic lepton, the SM charged leptons and $W^+$ (or to $\phi^+$ as shown in appendix C) is highly suppressed.

\subsection{Mass Splitting}

Although the masses of the triplet leptons are degenerate at the tree-level, electroweak quantum corrections
lift the degeneracy. 
The mass difference induced by one-loop of SM gauge bosons are calculated to be \cite{Ibe:2006de} 
\begin{eqnarray}
	M_{X^{--}} -M_{X^{-}} &=& \frac{\alpha_2 M}{4 \pi} \left( (3 \sin^2 \tw -1) f \left(\frac{M_Z}{M} \right)
	+f \left(\frac{M_W}{M} \right) \right) \\
  M_{X^{-}} -M_{X^{0}} &=& \frac{\alpha_2 M}{4 \pi} \left( (\sin^2 \tw +1) \ f \left(\frac{M_Z}{M} \right)
	-f \left(\frac{M_W}{M} \right) \right), 
	\label{mdifactual}
\end{eqnarray}
where $M$ is the mass scale of the lepton triplet and
\begin{equation}
	f(r)= r \left[ 2 r^3 \log r -2r +(r^2 -4)^{1/2} (r^2 +2) \log \left( \frac{r^2 -2 -r \sqrt{r^2 -4}}{2} \right) \right],
\end{equation}
which gives 
\begin{eqnarray}
	M_{X^{--}} -M_{X^{-}} & \approx & 848~{\rm MeV}, \hspace{20pt}   M_{X^{-}} -M_{X^{0}} \approx 492 ~{\rm MeV},
  \label{Mdif}
\end{eqnarray}
with $3\%$ and $7\%$ of accuracy respectively in the whole range 200~GeV $< M < 1000$ GeV. 
Although these mass differences are crucial for determining the allowed decay modes of the model as to be discussed 
in detail, they are still very small compared to the mass scale itself. Thus for most practical purposes of the LHC analyses, 
we have
\begin{eqnarray}
	M = M_1 \approx M_2 \approx M_3.
\end{eqnarray}
The relations for the mixing elements are approximated by
\begin{eqnarray}
	V_{i4} \approx \left( 1 + \rd (\delta_{i4} - 1) \right) v^*_{E4} v_{Ni},\qquad
	V_{4i}^* \approx \left(1+ \sqrt{2} (\delta_{i4} - 1 )\right) v_{N4}^* v_{Ei}. 
	\label{newcon2} 
\end{eqnarray}
Thus, the only dimensionful parameter is $M$ and it is taken in the range of 200~GeV $< M < 1000$ GeV henceforth \footnote{The current lower bound on a generic charged lepton is $100.8$ GeV \cite{datagroup}. Our choice of the lower mass value is motivated by LHC sensitivity with an intregated luminosity of 1~fb$^{-1}$ as seen in section V.C.}. Moreover, since $v_{Ni}$ and $v_{Ei}$ are related 
to each other by Eq. (\ref{vdefap}), we can take $v_{Ei}$ as the only independent couplings of the heavy leptons to the SM particles. 
\subsection{Lepton Interactions and PMNS matrix}

\begin{table}[tb]
	\centering
		\begin{tabular}{|c|c|c|c|c|} \hline
			$\psi_1$ & $\psi_2$ & Boson & $g_V$ & $g_A$
\\\hline
			$\nu_i$ &  $\nu_j$ & $Z^0$   
			& $ \frac{1}{4}( \delta_{ij} + v_{Ni} v_{Nj}^*)$      
			& The same
\\\hline
			$\nu_i$ &  $X^0$ & $Z^0$  
			& $\frac{1}{4} v_{Ni} v_{N4}^*$      
			& The same
\\\hline
			$X^0$ &  $X^0$ & $Z^0$  
			& 1
			& 0
\\\hline
			$e_i$ &  $e_j$ & $Z^0$  
			& $(\sin^2 \tw -\frac{1}{4})  \delta_{ij} +\frac{1}{4} v_{Ei} v_{Ej}^*$      
			& $-\frac{1}{4} \delta_{ij} +\frac{1}{4}  v_{Ei} v_{Ej}^*$      
\\\hline
			$e_i$ &  $X^-$ & $Z^0$  
			& $\frac{1}{4}  v_{Ei} v_{E4}^*$      
			& The same
\\\hline
			$X^-$ &  $X^-$ & $Z^0$  
			& $\sin^2 \tw$
			& $0$
\\\hline
			$X^{--}$ &  $X^{--}$ & $Z^0$  
			& $-1+2 \sin^2 \tw$    
			& 0
\\\hline
			$\nu_i$ &  $e_j$ & $W^{+}$
			& $\frac{1}{2} \left( V_{ij} + (\sqrt{2} -1) v_{Ni} v_{Ej}^* \right) $      
			& The same
\\\hline
			$\nu_i$ &  $X^-$ & $W^{+}$
			& $\frac{1}{2\sqrt{2}} v_{Ni} v_{E4}^*$      
			& The same
\\\hline
			$X^0$ &  $e_j$ & $W^{+}$
			& $0$     
			& $0$ 
\\\hline		
			$X^0$ &  $X^-$ & $W^{+}$
			& $\sqrt{2} \frac{v_{N4}}{v_{E4}}$     
			& $0$     
\\\hline
			$e_i$ &  $X^{--}$ & $W^{+}$
			& $\rd v_{Ei}$     
			& The same     
\\\hline
			$X^-$ &  $X^{--}$ & $W^{+}$
			& $\sqrt{2} v_{E4}$     
			& $0$     
\\\hline
		\end{tabular}
	\caption{ \footnotesize Couplings of the gauge bosons to the leptons in the mass eigenstates basis, as parameterized in a
	Lagrangian in Eq.~(\ref{Lmod}).} 
	\label{T1}
\end{table}
We now specify the lepton interactions with the SM gauge bosons. 
In the basis of mass eigenstates, we parameterize the coupling by a 
Lagrangian\footnote{To fix our normalization, $\tilde{g}$ is ${g}/{\sqrt{2}}~({g}/{\cos \tw})$ for $V=W^{\pm}~(Z^{0})$ 
in the SM.}
\begin{equation}
	{\cal L}  =  \tilde{g} \overline{\psi_1} \gamma_\mu (g_V - g_A \gamma_5) \psi_2 V^ \mu.
	\label{Lmod}
\end{equation}
The results are compiled in table \ref{T1}.  The details of the construction are in appendix B. It is noted that the coupling among $X^0$, the standard model charged leptons and $W^+$ is not zero but proportional to $m_{e_i}/M$, and hence highly suppressed. 

It is easy to see that by setting $v_{E}=v_{N}=(0,0,0,1)$, the SM couplings of the gauge fields to the leptons are recovered. Therefore, all the new physics involving the exotic leptons and the SM particles is encoded in $v_E$ and $v_N$, as introduced before. 
The $V-A$ structure of the charged weak interactions among the SM leptons is not modified by the presence of the exotic leptons. 

The corresponding $3\otimes 3$ PMNS matrix within this model, according to table \ref{T1}, is given by
\begin{eqnarray}
	U_{\alpha i}=V_{i \alpha}^* + (\text{\footnotesize$\sqrt{2}$} -1)v_{E \alpha}  v_{N i}^* \hspace{15pt} \alpha = e, \mu, \tau; ~i =1,2,3,
	\label{PMNS1}
\end{eqnarray}
or using (\ref{vdefap})
\begin{eqnarray}
	U_{\alpha i} = \left[ \left(1 + (\text{\footnotesize$\sqrt{2}$} -1) v_E v_E^\dagger \right)V^\dagger \right]_{\alpha i} = \left[V^\dagger  \left(1 + (\text{\footnotesize$\sqrt{2}$} -1) v_N v_N^\dagger \right)\right]_{\alpha i},
	\label{PMNSc}
\end{eqnarray}
This clearly shows that this matrix is not unitary. In fact equations (\ref{newcon2}) show that
\begin{eqnarray}
	\sum_{\alpha=e,\mu,\tau}  U_{\alpha i}^*U_{\alpha j} = \delta_{ij} + \frac{1}{2}v_{Ni} v_{Nj}^* \qquad
	\sum_{i=1}^3  U_{\alpha i}U_{\beta i}^* = \delta_{\alpha\beta} + v_{E\alpha} v_{E \beta}^*.
	\label{PMNS2}
\end{eqnarray}
Furthermore, using  the relations~(\ref{PMNS1}) and~(\ref{PMNS2}) , it can be shown that
\begin{equation}
	v_{Ni}= \sqrt{2} \sum_{\alpha=e,\mu,\tau} U_{\alpha i}^* v_{E\alpha}.
	\label{vNTovE}
\end{equation}
Finally, according to (\ref{Ly}), the interaction of the mass eigenstate leptons in the model with the Higgs is given in the unitary gauge by
\begin{eqnarray}
	-{\cal L}_H &=& \frac{1}{v} h(x) \overline{E_L} M_{Ed} \left( 1- v_E v_E ^\dagger \right) E_R 
	  + \frac{1}{v} h(x) \overline{N_L} M_{Nd} \left( 1- v_N v_N ^\dagger \right) N_R + h.c.
	 \label{Higgs}
\end{eqnarray}
Notice that there are non-diagonal terms that allow the Higgs boson to decay in SM leptons of different flavor, which is forbidden in the standard model at tree level. The couplings for an arbitrary gauge are given in appendix C.

\section{Current constraints on the model parameters}

\subsection{FCNC decays}

The new leptons contribute to flavor-changing processes. The absence of such decays put stringent bounds on the new particle and interactions. The particle data group \cite{datagroup} has compiled the constraints on these rare processes, which we report on table \ref{Geee}.
\begin{table}[t]
	\centering
	\begin{tabular}{|c|c|} \hline
		    Process  & $Br <$ \\\hline
$\mu^- \to e^- \gamma$ & $1.2 \times 10^{-11}$\\
$\mu^- \to e^-  e^-  e^+$ & $1.0 \times 10^{-12}$\\
$\tau^- \to e^- \gamma$ & $3.3 \times 10^{-8}$\\
$\tau^- \to \mu^- \gamma$ & $4.4 \times 10^{-8}$\\
$\tau^- \to e^- \mu^+ \mu^-$ & $2.7 \times 10^{-8}$\\
$\tau^- \to \mu^- \mu^+ \mu^-$ & $2.1 \times 10^{-8}$\\
$\tau^- \to e^- e^+ e^-$ & $2.7 \times 10^{-8}$\\
$\tau^- \to \mu^- e^+ e^-$ & $1.8 \times 10^{-8}$\\\hline
	\end{tabular}
	\caption{\footnotesize Branching fraction current upper limit for FCNC processes induced by exotic leptons.}
	\label{Geee}
\end{table}
We now derive the bounds on the couplings.

\subsubsection{$l_2 \to l_1  \gamma$}  

We evaluate the one-loop contribution from the new leptons to this process,
which is presented in detail in Appendix D. In contrast to models with only heavy electron-like leptons or heavy neutrinos, in our model we must also consider the contribution of the doubly-charged leptons. However, our results are consistent with previous studies ~\cite{Chua:2010me,Abada:2008ea,chengli}. As it is shown in appendix D, due to the relations as in Eq.~(\ref{newcon2}), the diagrams with the exotic neutral lepton are highly suppressed, and therefore the leading contributions come from diagrams with charged leptons. Also previous work \cite{Couture:1997eq,Wilczek:1977wb} includes contributions from doubly charged leptons. However those do not include heavy neutral or electron-like leptons along with the doubly charged lepton in the loops. As a result, their calculation is qualitatively different from the corresponding one for a triplet with hypercharge $Y=1$. After a careful calculation, we find the branching fraction to be given by
\begin{eqnarray}
\hspace{-1cm}
	{\rm Br}( l_2 \to l_1 \gamma) &=& \left( - 8 + 12 \sin^2 \theta_W + 8 g(r_W)+ \left(2 +\frac{1}{r_Z} \right) f(r_Z) + 8 \left(2 +\frac{1}{r_W} \right) f(r_W) + \frac{f(r_H) }{r_H}  	\right)^2 \nonumber\\ 
&&\times \left( \frac{G_F^2 m_{l_2}^5}{192 \pi^3 \Gamma_{l_2}} \right) \left(\frac{3 \alpha}{32 \pi}\right)  |v_{l_1}|^2 |v_{l_2}|^2 ,
\end{eqnarray}
where $r_a= {M_a^2}/{M^2}$, $\Gamma_{l_2}$ is the total width of the lepton $l_2$ and the functions $f$ and $g$ are defined in appendix D.

\begin{figure}[tb]
	\includegraphics[scale=1, width=8cm]{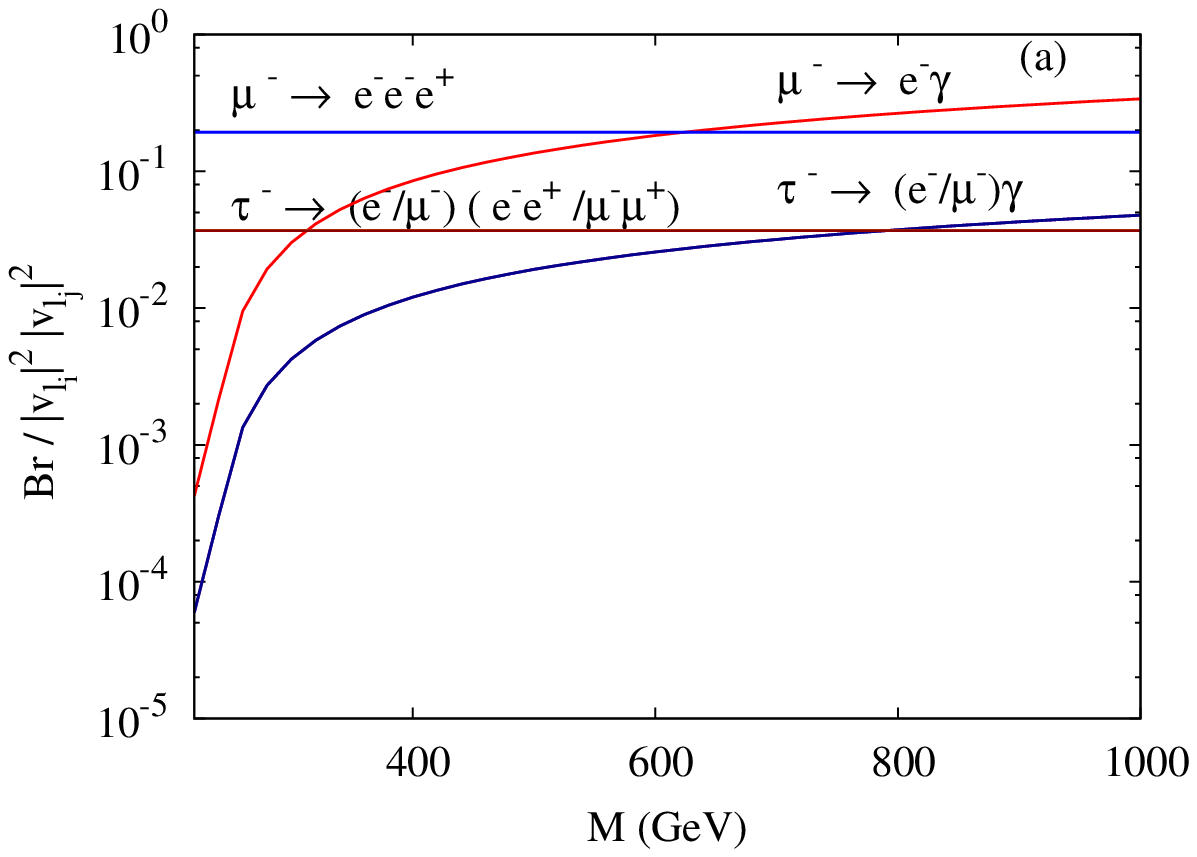} 	
	\includegraphics[scale=1, width=8cm]{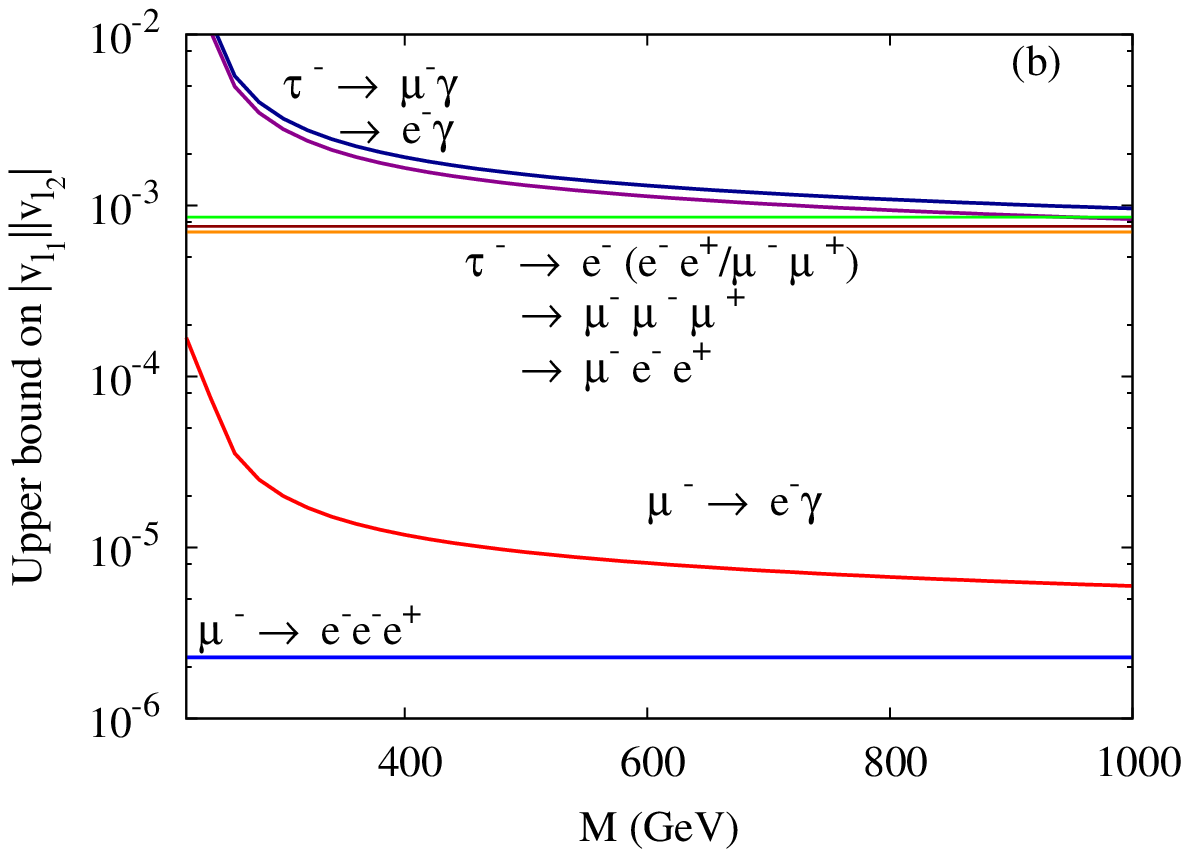} 	
	\caption{(a) Branching fraction of $l_2 \to l_1 \gamma$ and $l_2 \to l_1 l^- l^+$ as function of the triplet mass. 
	 (b) Upper bound on $|v_{l_1}||v_{l_2}|$ according to the experimental constraint in table \ref{Geee}.}
	\label{brmu}
\end{figure}

We present the branching fraction for this process as function of the triplet mass in Fig.~\ref{brmu}(a) after removing the
mixing parameters $ |v_{l_1}|^2 |v_{l_2}|^2$. 
Here and henceforth, we take $M_H=120$ GeV, but our results are not very sensitive to the particular value of the Higgs mass.
We can see that the branching fraction increases with the triplet mass logarithmically according to the asymptotic behavior of the functions $f$ and $g$. This is due to the enhanced coupling of the Higgs to the lepton triplet. 
We note that with a fully model consideration, the mixing parameters go like $ |v_{l_1} v_{l_2} | \sim 1/M^{2}$, and thus the physical
branching fraction asymptotically approaches zero at large mass, reflecting the decoupling behavior.
We translate the current bound of table \ref{Geee} to the mixing parameters as shown in Fig.~\ref{brmu}(b). 
As a result, setting $M = 1000$ GeV gives us the upper bound of the couplings as
\begin{eqnarray}
|v_e||v_\mu|<5.9\times10^{-6}, \;\;\;\;\; |v_e||v_\tau|<9.6\times10^{-4}, \;\;\;\;\;|v_\tau||v_\mu|<8.3\times10^{-4}.
\end{eqnarray}

\subsubsection{$l_2 \to l_1  l^- l^+ $} 

Due to the lepton-flavor changing interactions in this model, there are only two diagrams at tree-level, one involving a $Z$ boson, and the other one a Higgs boson. The one with the Higgs boson is however highly suppressed because it is proportional to the light fermion
masses like ${G_F m_{l_1} m_{l_2}}$. The corresponding branching fraction is found to be
\begin{eqnarray}
\hspace{-10pt}   {\rm Br}( l_2 \to l_1 l^- l^+) &=& \dfrac{\Gamma\left(l_2 \to l_1 l^- l^+\right)}{\Gamma_{l_2}}
    = \left( \dfrac{G_F^2 m_{l_2}^5}{192 \pi^3 \Gamma_{l_2}} \right) \left[ ( \sin^2 \tw)^2 +  2 ( \sin^2 \tw -\dfrac{1}{2} )^2  \ \right] \vert v_{l_2} \vert ^2 \vert v_{l_1}\vert ^2,
\end{eqnarray} 
Thus, table \ref{Geee} implies
\begin{eqnarray}
	|v_e||v_\mu|<2.2\times10^{-6}, \;\;\;\;\; |v_e||v_\tau|<8.6\times10^{-4}, \;\;\;\;\;|v_\tau||v_\mu|<7.0 \times10^{-4}.
\end{eqnarray}
The branching fraction and the bound are both indicated by the straight lines in Fig.~\ref{brmu}. It turns out that the numerical bound
on the couplings are slightly stronger for the $l_2 \to l_1 l^- l^+$ process.

\subsubsection{FCNC decays of $Z^0$}

Since the coupling of the $Z^0$ boson to SM leptons is not diagonal, they lead to Flavor Changing Neutral currents (FCNC) 
at tree level, and thus to the $Z^0$ boson might decay to SM leptons of different flavor. The decay rates for these processes are
\begin{eqnarray}
	\Gamma ( Z^0 \to \ell^-_i \ell^+_j) =
			\frac{G_F M_Z^3}{12\sqrt{2} \pi} \vert v_{Ei} \vert^2 \vert v_{Ej} 	\vert^2, \quad
	\Gamma ( Z^0 \to \nu_i \overline{\nu_j}) =
			\frac{G_F M_Z^3}{12\sqrt{2} \pi} \vert v_{Ni} \vert^2 \vert v_{Nj} 	\vert^2 \quad (\ i\neq j).
\end{eqnarray}
All these processes lead to constraints on the quantities $v_{Ei}$ and $v_{Ni}$, but not nearly as strong as those obtained 
from the $\mu$ rare decays above.

Finally, the exotic leptons generate contributions to the oblique corrections of
the gauge boson masses, the $S$ and $T$ parameters \cite{Peskin:1990zt}. It turns out that the constraints coming from FCNC constraints are more severe than any one coming from the EW precision parameters, so we will not pursue this study.

\subsection{non-unitarity of PMNS matrix}

It has been shown \cite{FernandezMartinez:2007ms} that if the PMNS matrix is not unitary and if it is written as $ U = (1+ \eta) U_0$ where $\eta$ is a hermitian matrix and $U_0$ a unitary matrix (which is always possible for an arbitrary matrix), then:
\begin{equation}
	|\eta| = \begin{pmatrix} |\eta_{ee}|  &  |\eta_{e \mu}| & |\eta_{e\tau}| \\. & |\eta_{\mu\mu} |& |\eta_{\mu\tau}| \\. & .& |\eta_{\tau\tau}| \end{pmatrix}  \lesssim \begin{pmatrix}  2.0\times10^{-3}  &  5.9\times10^{-5} & 1.6 \times10^{-3} \\ .& 8.2\times10^{-3} & 1.0 \times10^{-3}  \\.&.&2.6 \times10^{-3}   \end{pmatrix} 
\end{equation}
For our model, it is easy to show from equation (\ref{PMNS2}) that to leading order 
\begin{equation}
	\eta_{\alpha \beta} = \frac{1}{2}  |v_{E\alpha}||v_{E\beta}|
\end{equation}
which implies that:
\begin{eqnarray}
	|v_e||v_\mu|&<&1.2\times10^{-4}, \;\;\;\;\; |v_e||v_\tau|<3.2\times10^{-3}, \;\;\;\;\;|v_\tau||v_\mu|<2.0 \times10^{-3}, \nonumber\\
	|v_e|&<&6.3\times10^{-2}, \;\;\;\;\; |v_\mu|<1.3\times10^{-1},\;\;\;\;\;|v_\tau|<7.2 \times10^{-2}.
\end{eqnarray}
We can see that these constraints are not as strigent as those found in the previous section. However, now we have constraints on individual $v_i$.
\subsection{$\mu \to e$ conversion in heavy nuclei }

It is also possible to obtain a bound from $\mu \to e$ conversion in heavy nuclei. We will study this process for ${}^{48}_{22}Ti$ for which the current limit  \cite{datagroup} is
\begin{equation}
	R = \frac{\sigma(\mu^- Ti \to e^- Ti)}{\sigma(\mu^-Ti \to \text{capture})} < 4.3 \times 10^{-12}.
\end{equation}
Due to a $Z$ boson exchange, such process may take place in our model by means of the effective lagrangian:
\begin{equation}
 {\cal L}_{eff} =  -\frac{G_F v_{e}^{} v_{\mu}^* }{\sqrt{2}} \overline{e} \gamma^\lambda (1-\gamma^5) \mu \left(\overline{d} \gamma_\lambda\left( -\dfrac{1}{4}+\dfrac{1}{3} \sin^2 \theta_W + \dfrac{1}{4}\gamma^5\right) d + \overline{u}\gamma_\lambda\left(\dfrac{1}{4}-\frac{2}{3} \sin^2 \theta_W-\dfrac{1}{4}\gamma^5\right) u\right) .
\end{equation}
By using a standard formula, for example Eq. (2.16) of Ref. \cite{Bernabeu:1993ta}, we obtain $R =  0.992 |v_e|^2|v_\mu|^2$, which implies:
\begin{eqnarray}
	|v_e||v_\mu|&<& 2.1 \times10^{-6}. \nonumber
\end{eqnarray}
This constraint, although of the same order of magnitude, is more stringent than the one we got from $\mu \to eee$. However, it is subject to the theoretical and experimental uncertainties of nuclear physics.

\section{Decays of the Exotic Leptons}

\begin{table}[tb]
	\centering
\scalebox{1.2}{
		\begin{tabular}{|l|l|c|} \hline
		 & {\bf Channel} & Partial Width $ ~~\Gamma/ \frac{G_F M^3}{16\sqrt2\pi}$ \\\hline
\multirow{8}{*}{$X \to \ell$}
&$X^{--} \to e^-_i W^-$ & $ 4 F_1(r_W) \vert v_{Ei}\vert^2 $ \\\cline{2-3}
&$X^- \to e^-_i H $ & $F_0(r_H) \vert v_{Ei}\vert^2 $\\

&$X^- \to e^-_i Z^0 $ & $F_1(r_Z)\vert v_{Ei}\vert^2$ \\
&$X^- \to \nu_i W^-$ & $F_1(r_W)\vert v_{Ni}\vert^2$ \\\cline{2-3}
&$X^0 \to \nu_i H $ & $ F_0(r_H) \vert v_{Ni}\vert^2$\\
&$X^0 \to \nu_i Z^0 $ & $F_1(r_Z) \vert v_{Ni}\vert^2$\\
&$X^0 \to e^-_i W^+$ & $ 0 $ \\\hline
\multirow{2}{*}{$X_p \to X_k$}
&$X_p \to X_k W^{-*} \to X_k + e^-_i + \overline{\nu_j}$ & $~\quad$ Eq.~(\ref{Gxixk}) \\\cline{2-3}
&$X_p \to X_k W^{-*} \to X_k  \Pi^-  $  &  $32 \sqrt2 k G_F f_{\Pi}^2 \vert V \vert^2 \left(\frac{\Delta M}{M}\right)^3\sqrt{1-\left(\frac{ m_\Pi}{\Delta M}\right)^2}$ \\\hline
\end{tabular}
}
	\caption{Decay channels and partial widths for the exotic leptons, with $r_a = {m_a^2}/{M^2}<1$.
	$\Pi$ is a generic light meson ($\pi,\ K$)  and $f_\Pi$ its decay constant, $V$ is the corresponding CKM matrix element.
	$X_p$ generically denotes $X^{--}$ or $X^{-}$ and $X_k$ for $X^{-}$ or $X^0$, respectively. $k=6(1-\frac{4m_\pi^2}{m_\rho^2})^{-1}$ for the $\rho$ meson, otherwise $k=1$  \cite{Kumericki:2011hf}. }
	\label{tablerates}
\end{table}

To further study the phenomenology for the exotic leptons,  we now calculate the decays of the exotic leptons.
Depending on the masses, all the decay channels and the decay rate formulas have been listed in Table \ref{tablerates}. 

\subsection{Partial Decay Width}

\begin{figure}[tb]
	\includegraphics[scale=1, width=8cm]{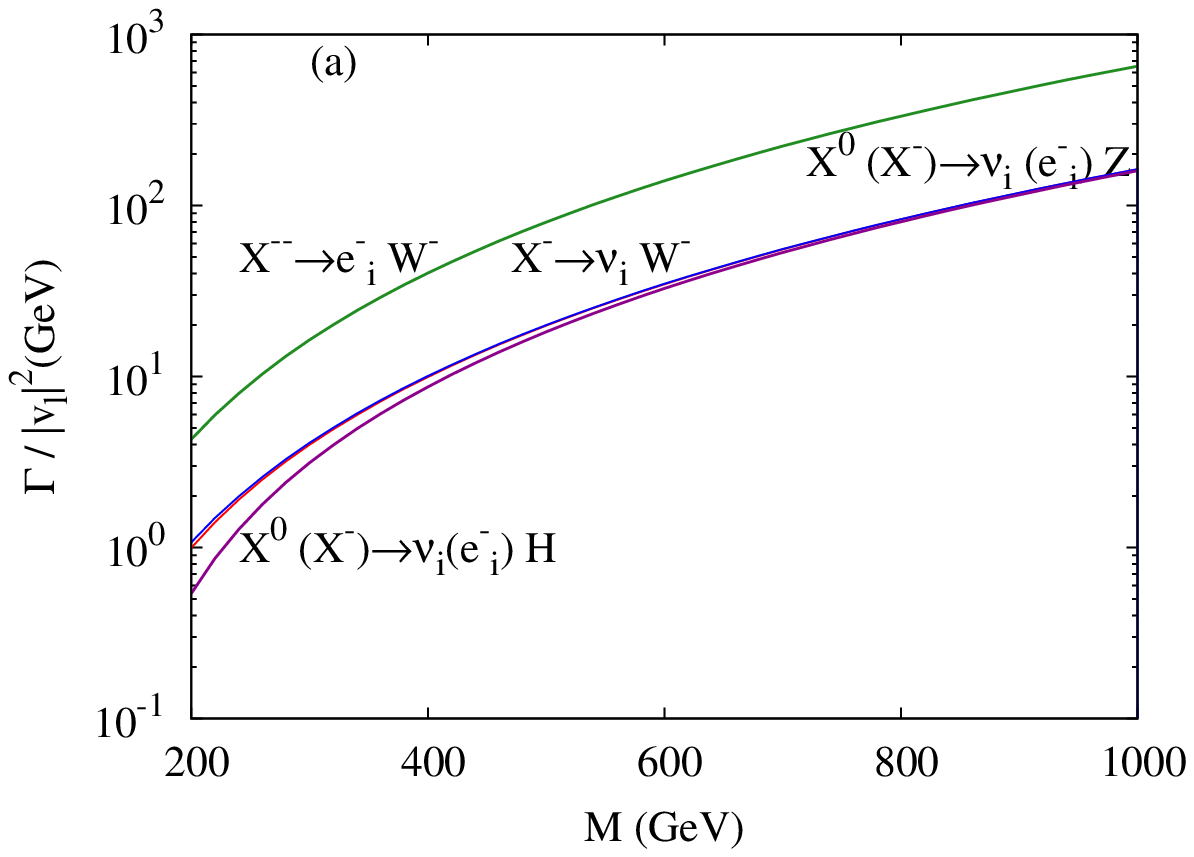} 	
	\includegraphics[scale=1, width=8cm]{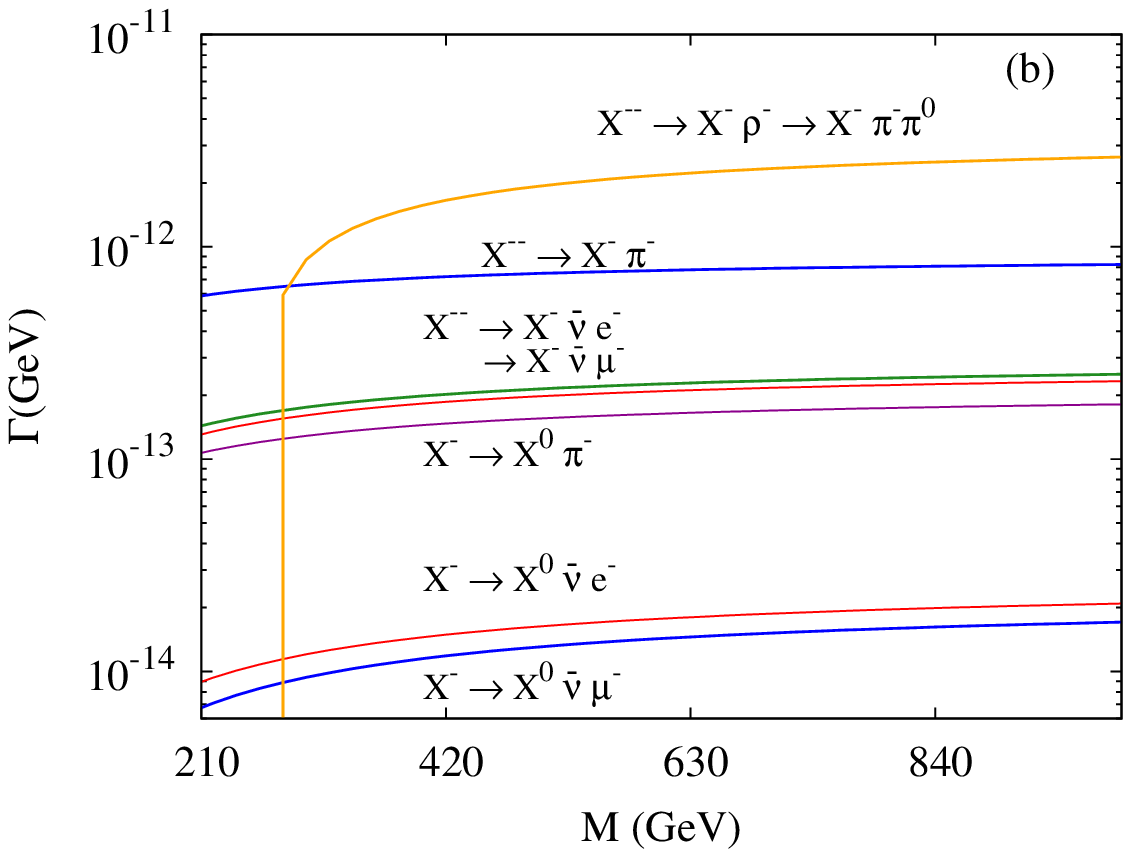} 	
	\caption{Decay rates (a) for $X \to \ell $ processes, with the mixing parameter $|v_\ell|^2$ factored out, where $\ell$ is the lepton in the final state; and (b) $X \to X$ processes, in the cases where the final state has neutrinos, we have summed over the three light states.}
	\label{Xto}
\end{figure}

For an exotic lepton above the scale of $M_{W}$, the important decay modes will be $X \to \ell+W,Z$ or $H$.
The decay width formulas are given in Table \ref{tablerates} as $X\to \ell$, with the functions
\begin{equation}
F_n(x) = (1-x)^2(1+2nx)^2. 
\label{Fns}
\end{equation}
As seen from those results, the partial widths are all proportional to the mixing angle squared between the heavy-light
transition. We plot the partial decay widths for those transitions versus the exotic lepton mass
in Fig.~\ref{Xto}(a). Once again we take $M_H=$120 GeV. The mixing angle squared has been
factored out for comparison. It is interesting to note that the decay width for $X^{\pm\pm}$ is about a factor of four larger than those of $X^{\pm}$ or $X^{0}$ , due to the gauge couplings in Table \ref{T1}. The similarity among the other channels is in accordance with the Goldstone boson equivalence theorem. 

The transition between two heavy states will have no mixing angle suppression ($|v_{E4}| \approx |v_{N4}| \approx 1$). However, it will suffer from the three-body phase space suppression due to the near mass degeneracy. 
For the leptonic final state $X_p \to X_k W^{-*} \to X_k + e^-_i + \overline{\nu_j}$, we have the expression, similar to the muon
decay 
\begin{eqnarray}
	\label{Gxixk}
	&&\frac{d\Gamma(X_p \to X_k  e^-_i  \overline{\nu_j})}{dx_k dx_i}= \frac{G_F^2M_p^5 \vert U_{ij} \vert^2}{8 \pi^3} \times 
	\nonumber\\ && 
	 \left( x_i (1-\mu_k + r_i - x_i)(-1+x_i +x_k -r_i - \mu_k)(2-x_i-x_k) -2 \sqrt{\mu_k} (1+\mu_k-r_i-x_k) \right), 
\end{eqnarray}
where $\mu_k = M_k^2/M_p^2,\ r_i = {m_i^2}/{M^2}$ and $x_a = {2 E_a}/{M}$. The integration ranges for the energy variables are
\begin{eqnarray}
    2 \sqrt{\mu_k} \leq x_k \leq 1+\mu_k-r_i, \ \ \ 
	x_i \lessgtr \frac{1}{2} \left[ (2-x_k) \left( \frac{1+\mu_k+r_i-x_k}{1+\mu_k-x_k} \right) \pm \sqrt{x^2_k-4\mu_k} \left(\frac{1+\mu_k- r_i-x_k}{1+\mu_k-x_k}\right) \right] .
	\nonumber
\end{eqnarray}
Notice that the mass difference between the exotic leptons, as given in Eq.~(\ref{mdifactual}), crucially controls the decay rates. Furthermore, 
since this difference is of the order of few hundred MeV, we have kept the charged lepton mass $m_i$ explicit in the calculation.

We plot the decay widths of $X_p \to X_k$ in Fig.~\ref{Xto}(b) for each exotic lepton versus $M$. We see that these rates vary very slowly with the exotic lepton mass. Furthermore it is important to realize that among these, the leading mode is the two-body decay into a pion for the singly-charged lepton or the two-body decay into a $\rho$ for a doubly-charged lepton when kinematically accessible. 
Using the formula given in Table~\ref{tablerates}, in the mass range $200< M<1000$ we have therefore, that
\begin{equation}
	\Gamma{(X^{--} \to X^-)} \approx 3.7 \times 10^{-12}~{\rm GeV}, \quad \Gamma{(X^- \to X^0)} \approx 2.3 \times 10^{-13}~{\rm GeV}.
\end{equation}

\subsection{Total Decay Widths And Branching Fractions}

When considering a total decay width summing over the contributing channels, we find it useful to introduce the notation
\begin{equation}
		\lambda = \sum_{i=e,\mu,\tau} \vert v_{i} \vert^2
		 \label{lambda},
\end{equation}
that controls the heavy-light ($X\to \ell$) transition. 
Since deviations of the PMNS matrix from unitarity are of second order in $v_{Ei}$, and because of 
Eq.~(\ref{vNTovE}), for the neutrino couplings we have
\begin{equation}
	 \sum_{i=1}^3 \vert v_{Ni} \vert^2 = 2 \lambda + {\cal O}(\lambda^2).
\end{equation}
Using this parameter, we express the total widths as 
\begin{eqnarray}
	\Gamma_{X^{--}} &=& \frac{G_F M^3}{4  \sqrt{2} \pi } F_1(r_W) \lambda + \Gamma(X^{--}\to X^{-}) 
	 \approx \left(\frac{M}{115 \ {\rm GeV}}\right)^3 \lambda +  \Gamma(X^{--}\to X^{-}) 
	    \label{totalG}\\
	\Gamma_{X^-} &=& \frac{G_F M^3}{16 \sqrt{2} \pi } \left( 2 F_1(r_W) + F_1(r_Z) +F_0(r_H) \right) \lambda +  \Gamma(X^{-}\to X^{0}) 
	\approx \left(\frac{M}{115\ {\rm GeV}}\right)^3 \lambda +  \Gamma(X^{-}\to X^{0})  \label{totalGap}\nonumber\\
	\Gamma_{X^0} &=& \frac{G_F M^3}{8 \sqrt{2} \pi } \left( F_1(r_Z) +F_0(r_H) \right) \lambda 
	 \approx  \left(\frac{M}{115\ {\rm GeV}}\right)^3 \lambda, \nonumber 
\end{eqnarray}
where $F_n$ was defined on Eq.~(\ref{Fns}) and the rates are in units of GeV.

The squared sum of the mixing angles $\lambda$ is of fundamental importance for the decay life time and branchings.
We recall that from the previous discussions, the experimental constraints discussed before put limits on the 
possible values of $\lambda \lesssim 10^{-6}$. 
We now categorize the phenomenology roughly according to the following two regions.

\begin{figure}[tb]
	\includegraphics[scale=0.8]{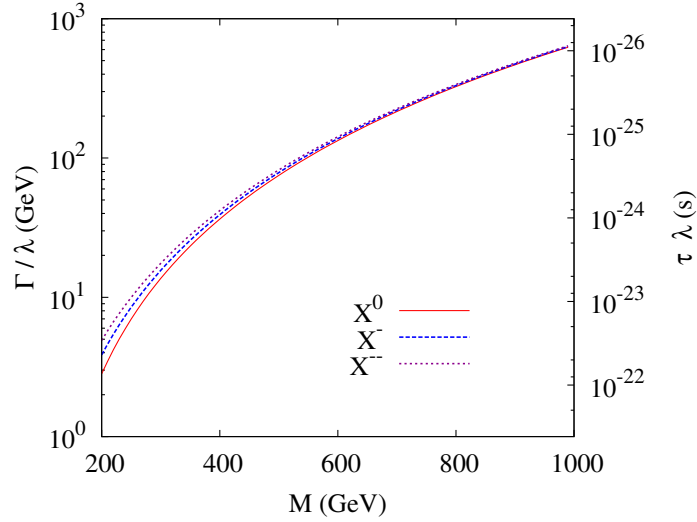} 
	\caption{Total widths (left axis) and lifetimes (right axis) as functions of exotic leptons mass for $10^{-12}<\lambda< 10^{-6}$.}
	\label{figbr1}
\end{figure}

\subsubsection{$10^{-12}<\lambda< 10^{-6}:\ X\to \ell$~ Transition Dominance}

 In this case the coupling of the exotic leptons to the standard model leptons is strong enough so that their leading decay modes are
 to the SM leptons ($X\to \ell$), along with gauge bosons or Higgs bosons. The total widths are proportional to $\lambda$. 
 These are plotted in Fig.~\ref{figbr1}, along with the the lifetime of each of the exotic leptons on the right-hand side axis. 
 The mixing angles squared are again factored out. 
 Taking into account these small mixings, the life time in this parameter region is still rather short, leading to prompt decays
 in collider experiments, although it may result in secondary vertices when $\lambda \sim 10^{-12}$.
 
\begin{figure}[t]
	\includegraphics[scale=1,width=8.1cm]{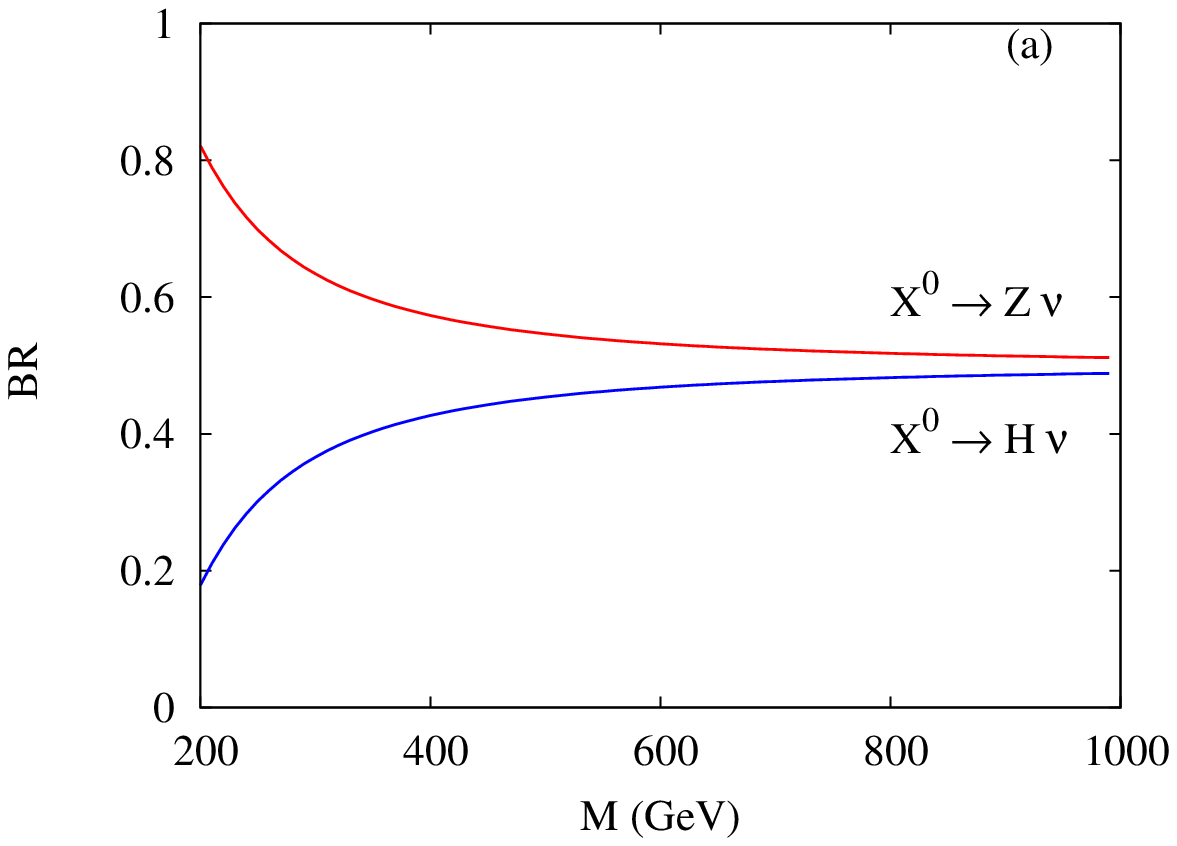} 
	\includegraphics[scale=1,width=8.1cm]{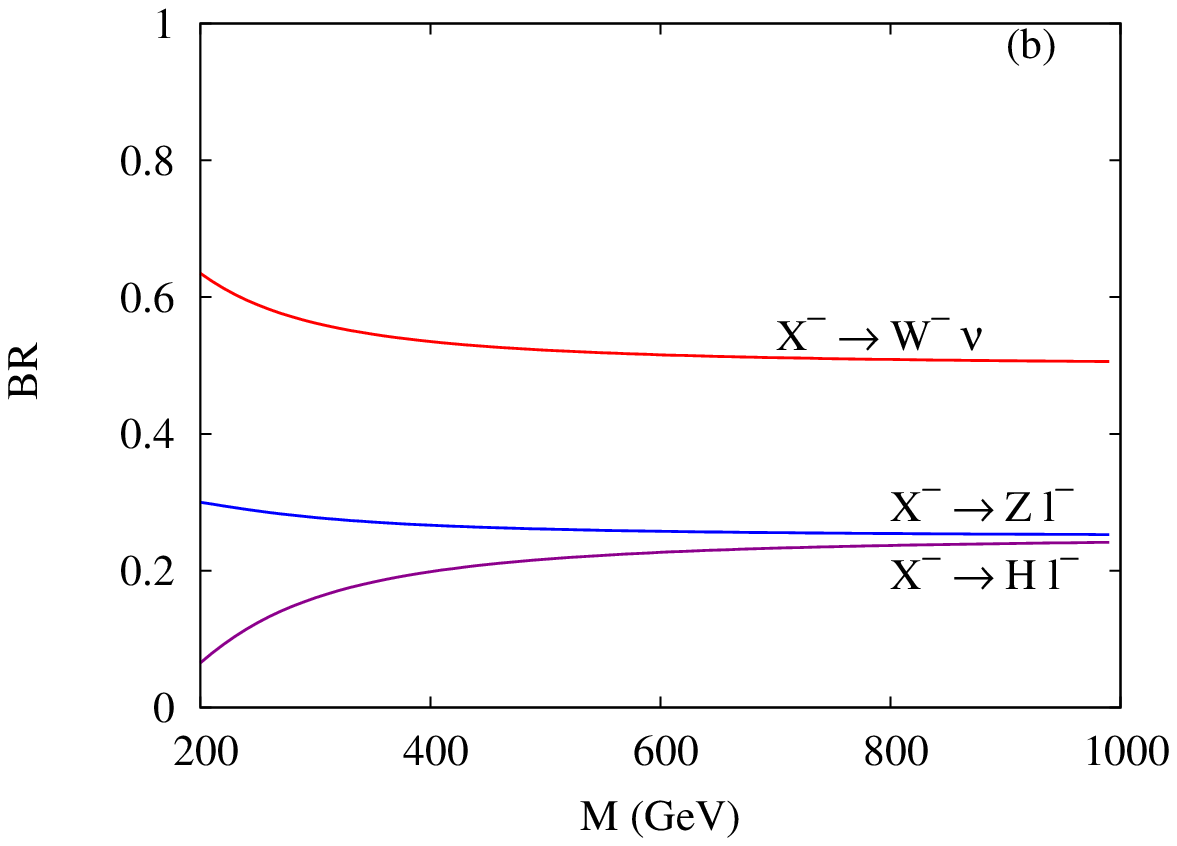} 
	\caption{Branching ratios for (a) $X^0$ and (b) $X^-$ for $10^{-12}<\lambda< 10^{-6}$ }
	\label{figbr2}
\end{figure}

 Figure \ref{figbr2} corresponds to the branching fractions. 
The SM lepton flavors are summed over as earlier. The branching fractions to $W,Z,H$ are again in accordance with the
Goldstone boson equivalence theorem at the high mass region. 
 $X^{--}$ decays to a charged lepton and a $W$ with a 100$\%$ branching fraction, 
The relative fraction to a specific charged
 lepton depends on the ratio of the mixing angles squared $|v_{e}|^{2} : |v_{\mu}|^{2} : |v_{\tau }|^{2} $.
 Determination of the leptonic branching fractions would lead to the most interesting phenomenology.
 
\subsubsection{$ \lambda< 10^{-13}:\ X_{p} \to X_{k}$~ Transition Dominance}
 
 In this case, 
 the leading decay mode of the charged exotic leptons is $ X^{\pm\pm} \to X^{\pm}\pi^\pm\pi^0 $ or $ X^{\pm} \to X^{0} \pi^\pm $. The lifetime is approximately constant,
 about the order $10^{-12}$ s. However, the nearly degenerate masses for $X$ make observable SM final state very
 soft, typically with an energy less than a GeV, and thus essentially escape from the detection in the collider environment. 
 The lightest exotic lepton, $X^{0}$, will only undergo a $X^{0} \to \nu$ transition, and thus difficult to detect as well 
in collider experiments. We will not consider this parameter range due to the lack of relevance for LHC phenomenology.


\section{Searches for the lepton triplet at the LHC}

\subsection{Total cross sections}

\begin{figure}[t]
	\includegraphics[scale=1, width=8cm]{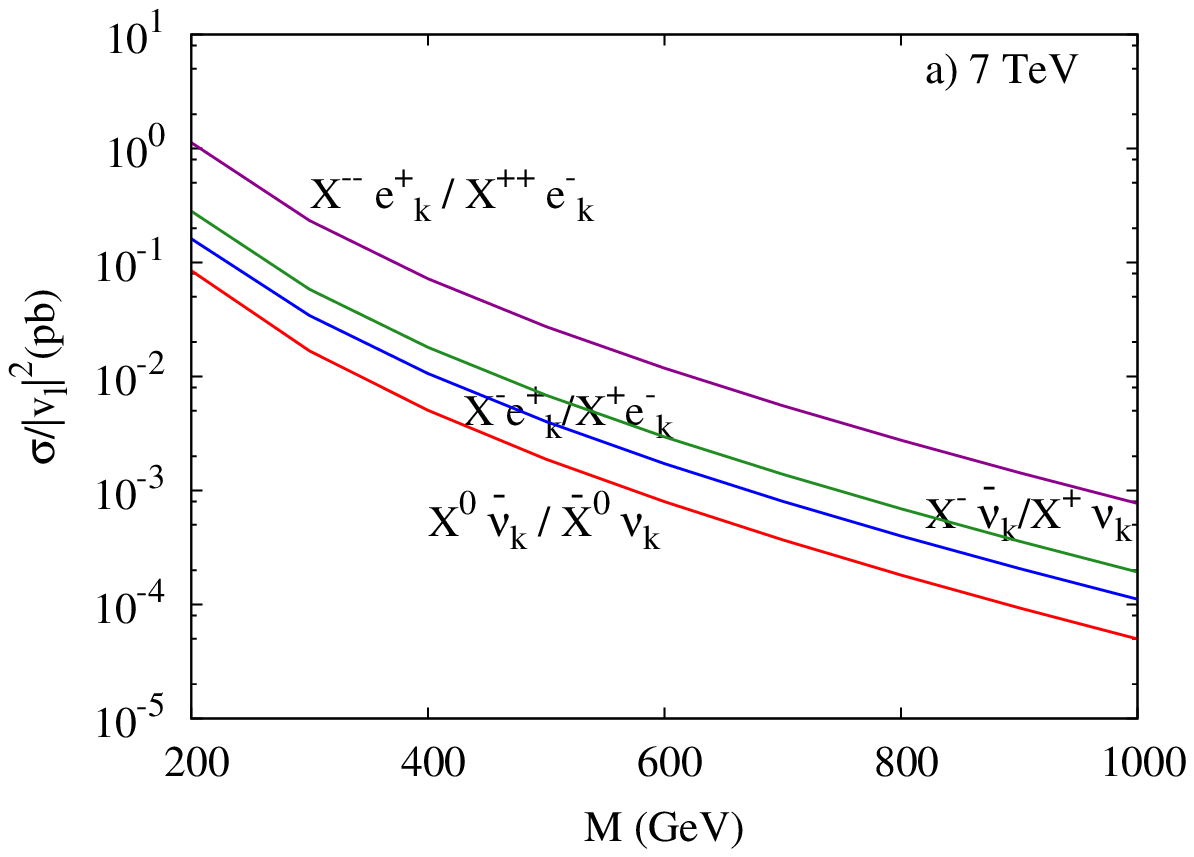} 
	\includegraphics[scale=1, width=8cm]{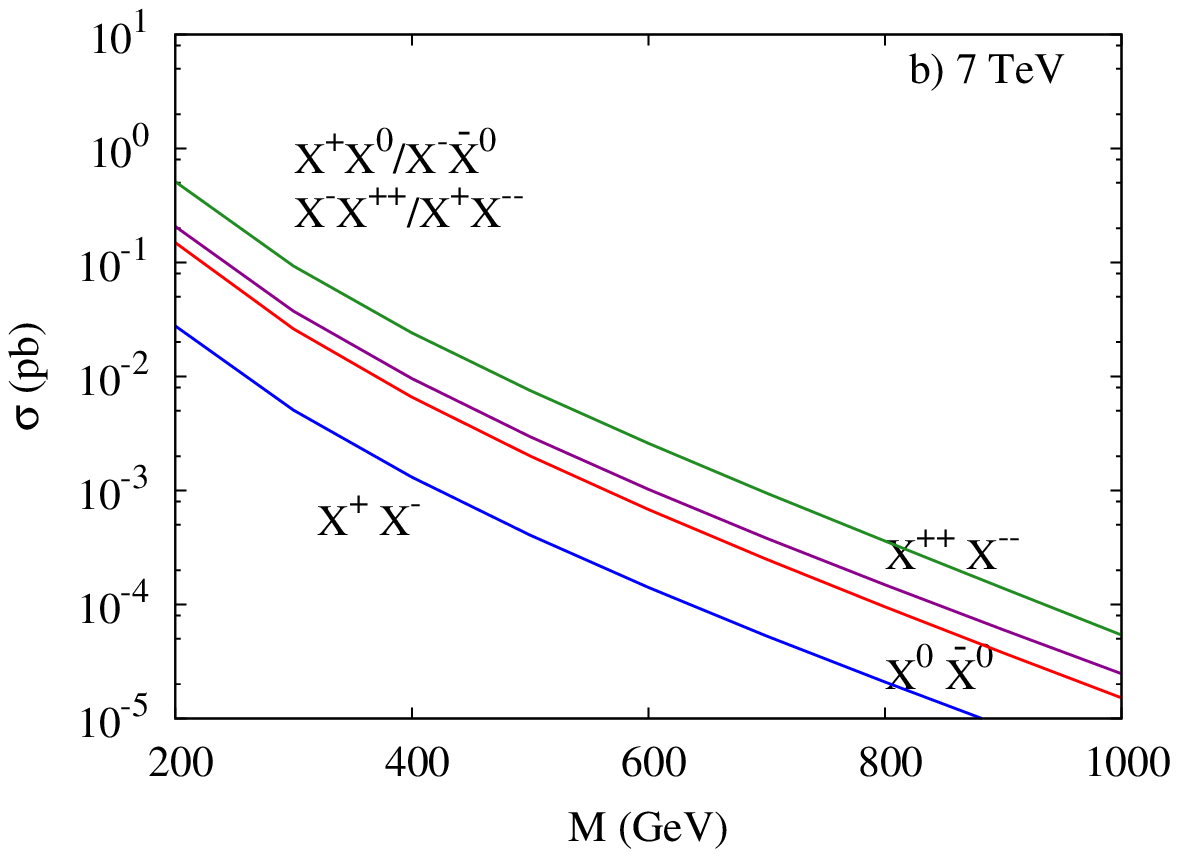} 
	\includegraphics[scale=1, width=8cm]{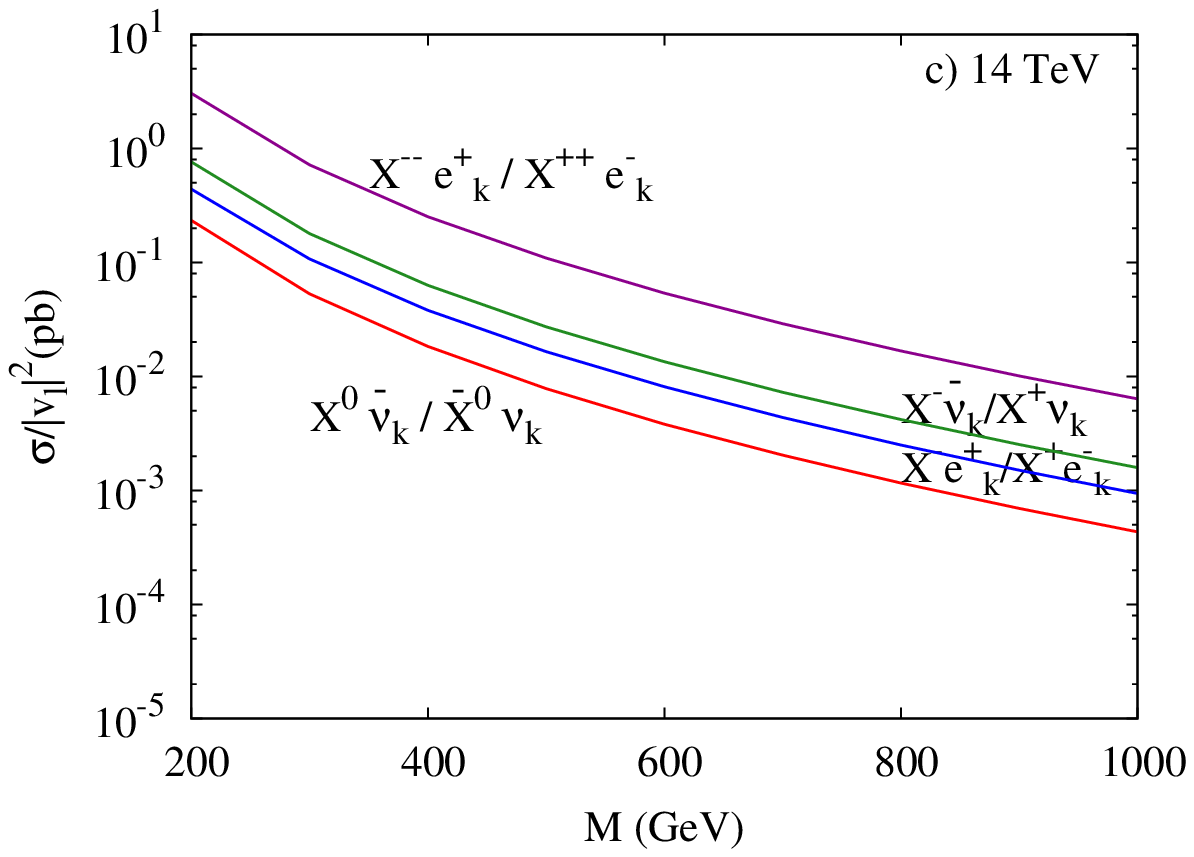} 
	\includegraphics[scale=1, width=8cm]{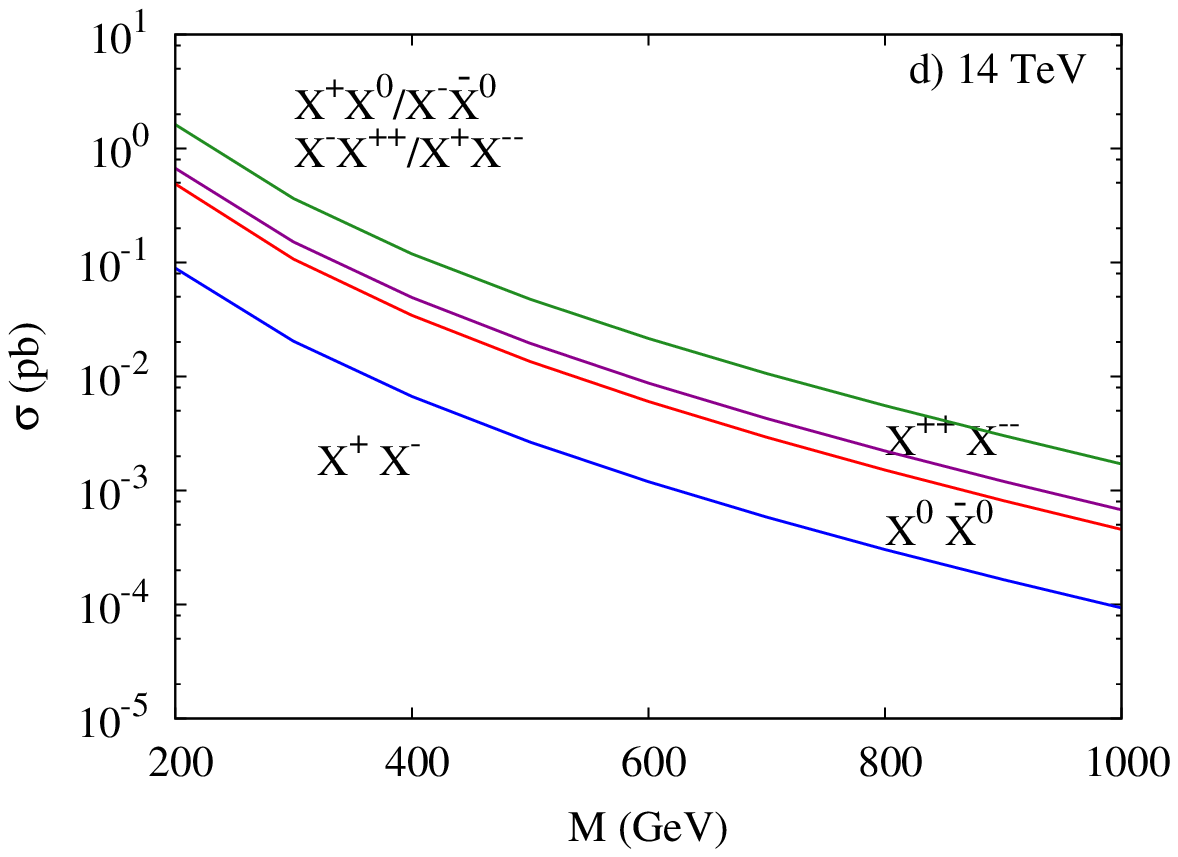} 
	\caption{Cross sections for the LHC (a) associated production at 7 TeV; (b) pair production at 7 TeV; (c) associated production at 14 TeV; (d) pair production at 14 TeV.}
	\label{sigmasf}
\end{figure}

We first present the total cross sections for all the possible processes for the exotic lepton production 
in Fig.~\ref{sigmasf} at the LHC for 7 TeV and 14 TeV. Once again, we factor out
the overall couplings. The associated production of an exotic lepton and a SM lepton 
is shown in Fig.~\ref{sigmasf}(a) and Fig.~\ref{sigmasf}(c). The production rates are
suppressed by the mixing angle squared. Since they are at least of the order of $10^{-6}$, their corresponding cross sections are negligible. 
We will thus only consider the pair production of the exotic leptons via the SM gauge interactions for their search at the LHC,
as shown in Fig.~\ref{sigmasf}(b) and Fig.~\ref{sigmasf}(d). 
Similar results are obtained in LHC searches of doubly charged Higgsinos ~\cite{higgsinos}.

\begin{table}[t]
	\centering
		\begin{tabular}{|c||c|c|c|c|c|c|} \hline
	              & $X^{++} \to \lp \wbp$ & $X^{+} \to \nub \wbp$ & $X^{+}  \to \lp Z$ & 
	                $X^{+} \to \lp H$ & $\overline{X^0} \to \nub Z$ & $\overline{X^0} \to \nub H$  
\\\hline
\hline
			$X^{--} \to \lm \wbm$&
			$ \lm \lp \wbm \wbp$&
			$ \lm \nub \wbm \wbp$&
			$ \lm \lp  \wbm Z$&
			$ \lm \lp \wbm  H $&
			-& 
			-

\\\hline
				$X^{-} \to \nu \wbm$&

			$\nu \lp \wbm \wbp $&
			$\nu \nub \wbm \wbp $&
			$\nu \lp \wbm Z $&
			$\nu \lp \wbm H $&
			$\nu \nub \wbm Z $&
			$\nu \nub \wbm H $
\\\hline
			$X^{-}  \to \lm Z$ &
			$\lm \lp Z \wbp$ &
			$\lm \nub Z \wbp$ &
			$\lm \lp Z Z$ &
			$\lm \lp Z H$ &
			$\lm \nub Z Z$ &
			$\lm \nub Z H$ 

\\\hline
			$X^{-} \to \lm H$ &
			$ \lm \lp H \wbp$ &
			$ \lm \nub H \wbp $ &
			$ \lm \lp H Z $ &
			$ \lm \lp H H$ &
			$ \lm \nub H Z $ &
			$ \lm \nub H H$ 

\\\hline
			$X^0 \to \nu Z$&
			-&
			$ \nu \nub Z \wbp$ &
			$ \nu \lp Z Z$ &
			$ \nu \lp Z H$ &
			$ \nu \nub Z Z$ &
			$ \nu \nub Z H$ 

\\\hline
			$X^0 \to \nu H$&
			-&
			$ \nu \nub H \wbp$ &
			$ \nu \lp H Z$ &
			$ \nu \lp H H$ &
			$ \nu \nub H Z$ &
			$ \nu \nub H H$ 
\\\hline
		\end{tabular}
	\caption{\footnotesize Exotic lepton decay channels to SM particles.} 
	\label{channels}
\end{table}

\subsection{Characteristic Final states for the Exotic Leptons}

We have seen that the pair production of the exotic leptons via the SM gauge interaction may lead to sizable rate.
In Table \ref{channels}, we list all the decay channels with a SM lepton in the final state. 
In order to test this model, it is necessary to identify the most characteristic feature of the model.
First, we would like to reconstruct the exotic lepton mass to claim a signal observation. Second, we wish to establish the
nature of the doubly charged lepton to be conclusive for the model. Third, we hope to choose a channel that keeps a large
signal rate while that stands out above the SM backgrounds. With these considerations, we focus out study in the following to
the production and decay modes
 \begin{equation}
pp \to X^{--} X^{++} \to \lm \wbm \ \  \lp \wbp.
\label{mode}
 \end{equation}
For simplicity from the observational point of view, we assume that $| v_e |\approx |v_{\mu}| \approx |v_{\tau} |$, and thus
 \begin{equation}
	BR(X^{\pm \pm} \to e^\pm W^\pm) \approx 	BR(X^{\pm \pm} \to \mu^\pm W^\pm) \approx 
		\frac{1}{3} \sum_i BR(X^{\pm \pm} \to \ell_i^\pm W^\pm) \approx \frac{1}{3}.
\end{equation}
We will consider only $e$ and $\mu$ final states for the sake of experimental identification. Furthermore, one of the $W$'s in the final state
is required to decay leptonically for the charge identification and the other $W$ to decay hadronically for the mass reconstruction. 
%
%
As a result, the final state for the channel in Eq.~(\ref{mode}) is
 \begin{equation}
 X^{--} X^{++} \to \lm \wbm \ \  \lp \wbp \to \ell^{-}\ell^{-}\nu\ \ \ell^{+} jj + {\rm h.c.},
\label{final}
 \end{equation}
with a total branching fraction
\begin{equation}
	 {\rm BR} \approx \frac{2}{3} \cdot \frac{2}{3} \cdot (0.676) \cdot (2 \cdot\ 0.107) \cdot 2 \approx 13\% . 
	 \label{BRjjlll}
\end{equation}

\subsection{Observability of Exotic Leptons at the LHC}

\subsubsection{Signal event selection}

We first define the signal identification. For definiteness, in this section we will specify the channel:
one positively charged lepton, two negatively charged leptons and two jets, plus missing energy. 
Following the detector coverage for the LHC experiments, we apply the following basic kinematical acceptance 
on the transverse momentum, rapidity, missing transverse energy, and the particle separation
\begin{eqnarray}
	p_T(\ell) &>& 15 ~\mbox{GeV}, ~ |\eta_\ell |< 2.5,\ {\not}E_T > 25 \mbox{ GeV} \\
	p_T(j) &>& 15 ~\mbox{GeV}, ~ |\eta_j |< 2.5 \\
	\Delta R(jj) &>& 0.4, ~ \Delta R(j\ell) > 0.4, ~ \Delta R(\ell \ell) > 0.3,
	\label{cutsLHC}
\end{eqnarray} 
where the particle separation is $\Delta R(\alpha \beta) \equiv \sqrt{(\Delta \phi_{\alpha \beta})^2+(\Delta \eta_{\alpha \beta})^2}$ with $ \Delta \phi$ and $\Delta \eta$ being the azimuthal angular separation and rapidity difference between two particles. 
To further simulate the detector effects, we
assume that the lepton and jet energies are smeared with a Gaussian distribution according to
\begin{equation}
	\frac{\delta E}{E} = \frac{a}{\sqrt{E/\mbox{GeV}}}\oplus b,
	\label{smearing}
\end{equation}
\noindent
where $a_\ell = 5\%$, $a_j = 100\%$, $b_\ell = 0.55\%$ and $b_j = 5\%$ \cite{Ball:2007zza}. 
We further require the hadronic $W$ reconstruction, taking the invariant mass of the jets in the range
\begin{equation}
	M_W-20\mbox{ GeV}<M_{jj}<M_W +20\mbox{ GeV}.
\end{equation}
We then calculate the invariant mass of the jets and the positively-charged lepton $M_{jj\ell^+}$, which we expect to yield the mass of $X^{++}$ for a signal. 
With the remaining two negatively-charged leptons, using the transverse momenta, we solve for all the possible neutrino momenta that would come from a $W$ boson decay. This gives four possible solutions. Since we are expecting $M_{\ell^-\ell^-\nub}$ to also yield the mass of $X^{--}$, we choose the solution that gives $M_{\ell^-\ell^-\nub}$ closest to $M_{jjl^+}$. 
We find that our reconstruction scheme quite efficient, with only about 3\% of events not leading to a solution.

\begin{figure}[tb]
	\includegraphics[scale=1, width=8cm]{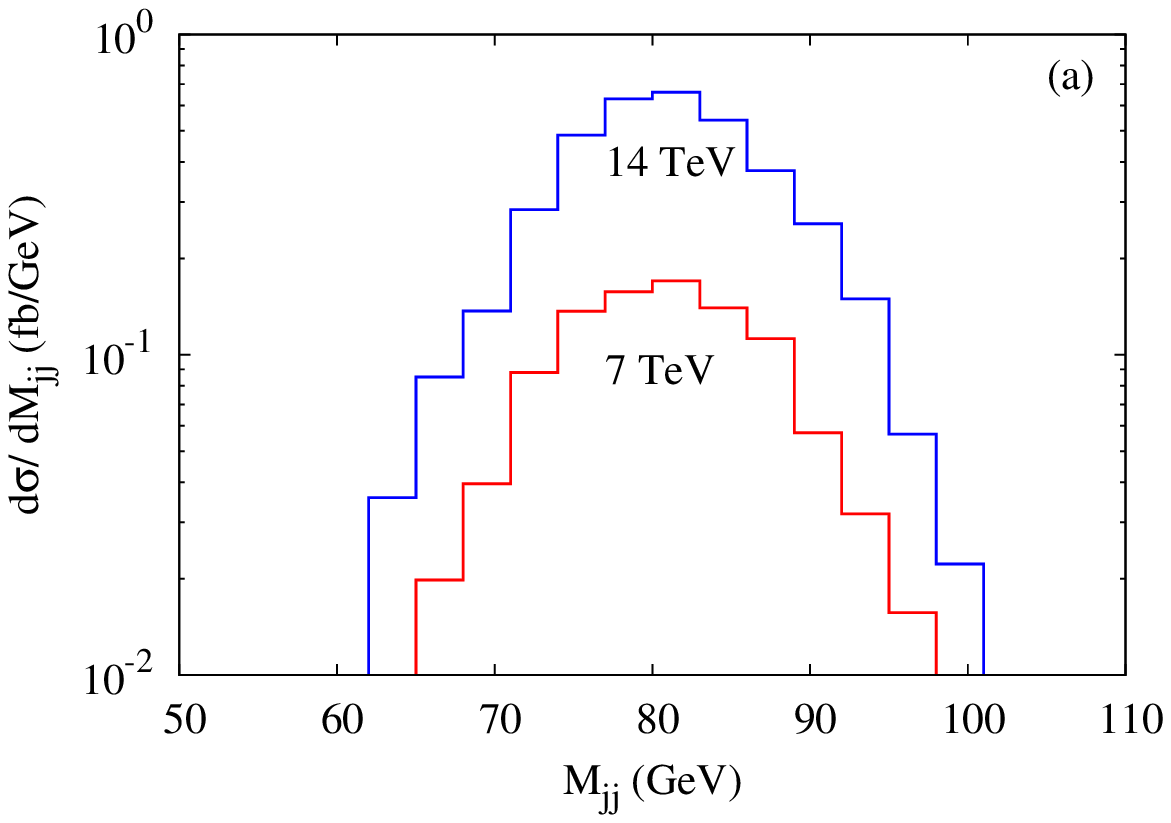} 
	\includegraphics[scale=1, width=8cm]{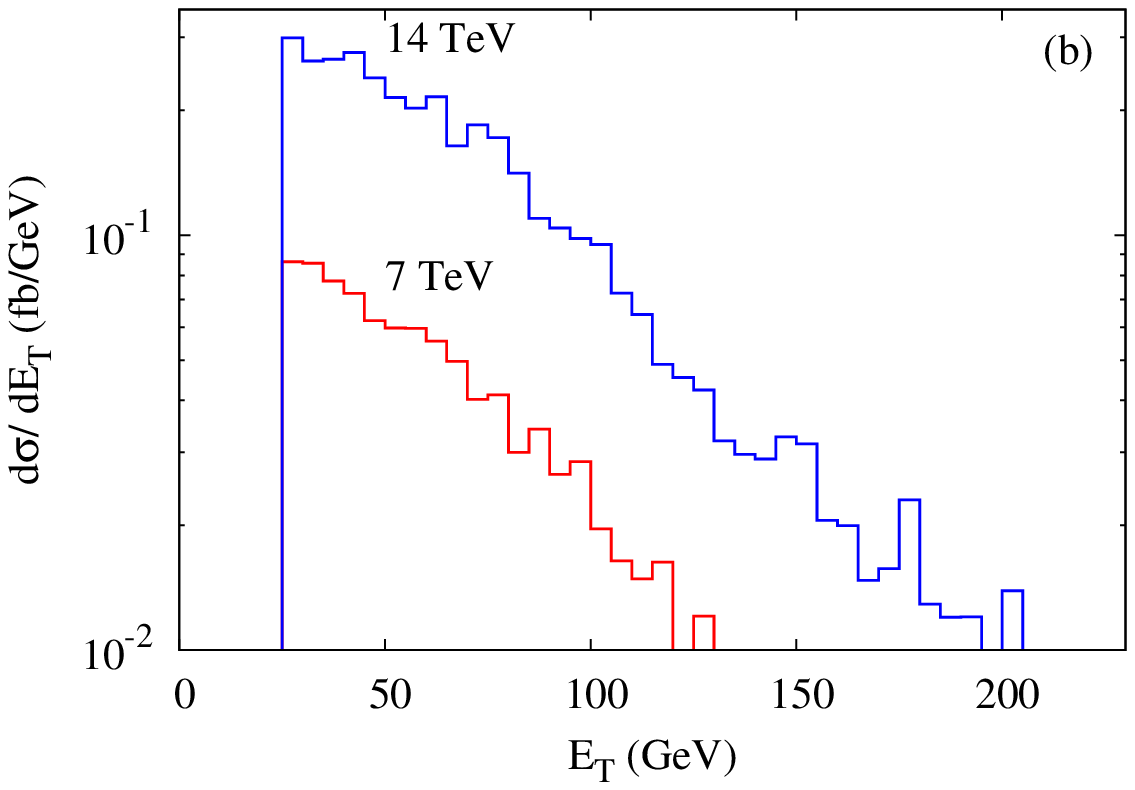} 
	\includegraphics[scale=1, width=8cm]{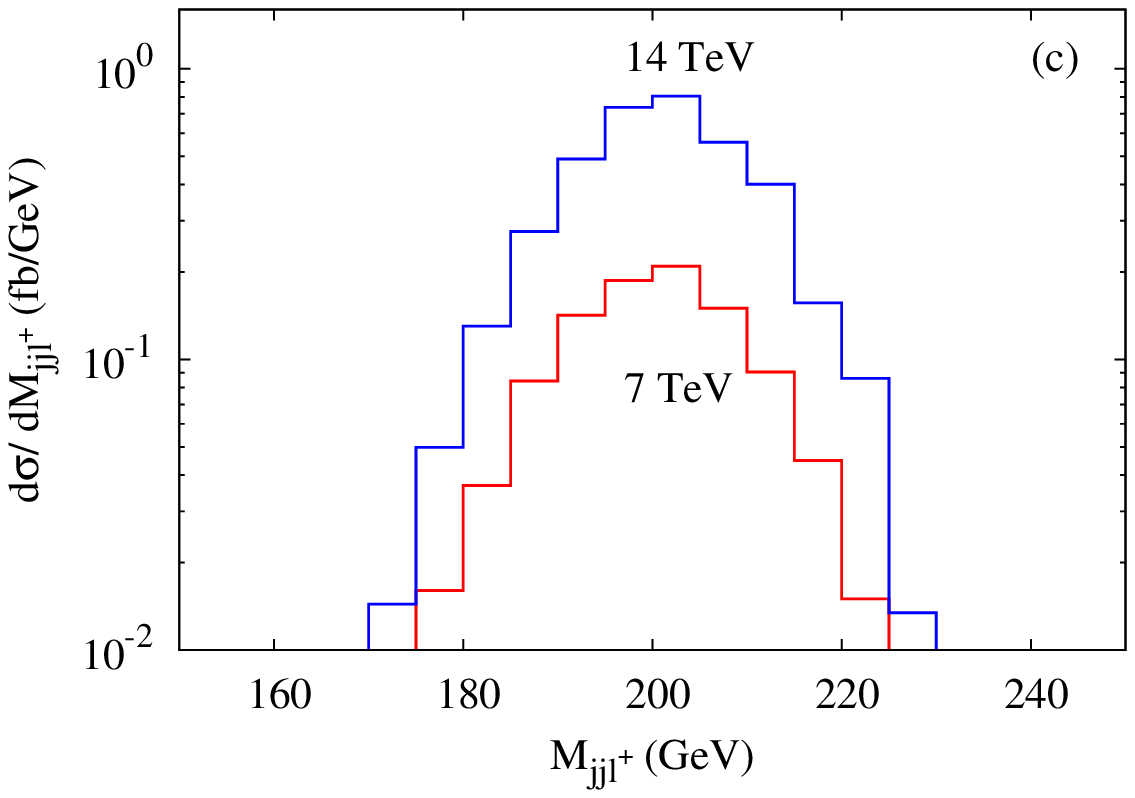} 
	\includegraphics[scale=1, width=8cm]{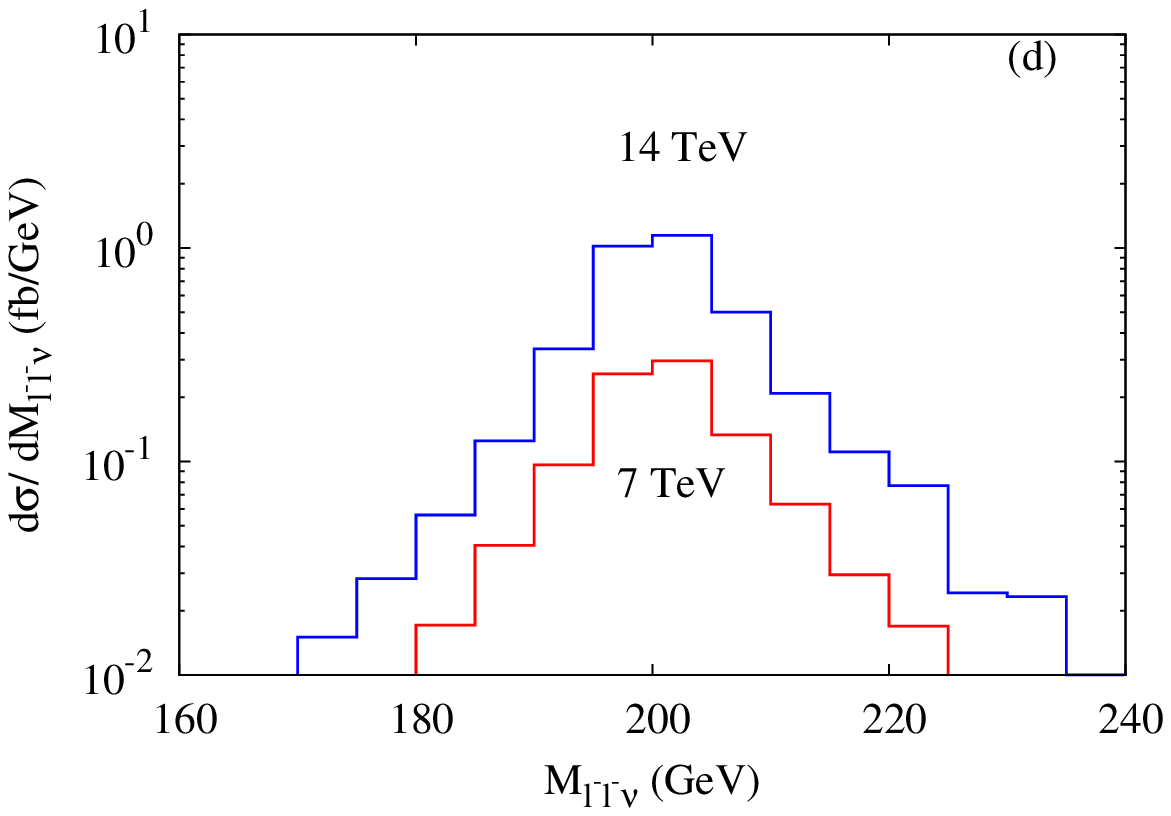} 
	\caption{Differential cross sections after reconstruction for the LHC at 7 TeV (lower curves) and 14 TeV (upper curves).}
   \label{diffcross}
\end{figure}

For the sake of illustration, we take the triplet mass to be 200 GeV, and perform the simulation for the LHC at 7 TeV and 14 TeV.
The differential cross sections are shown 
for the reconstructed $M_{W}$ in Fig.~\ref{diffcross}(a),
the missing transverse energy in Fig.~\ref{diffcross}(b),
the reconstructed exotic lepton mass in the hadronic $W$ mode in Fig.~\ref{diffcross}(c),
and the reconstructed exotic lepton mass in the leptonic $W$ mode in Fig.~\ref{diffcross}(d).
Finally, to strengthen the signal observation, it is possible to device a mass cut
\begin{eqnarray}
	M_{jj\ell^+}-30\mbox{ GeV}<M_{\ell^-\ell^-\nub}<M_{jjl^+}+30\mbox{ GeV},
	\label{mjll}
\end{eqnarray}
which does not affect our signal construction appreciably.

Although the final state under consideration is very clean and unique, there are still
some SM backgrounds that lead to similar final states to our signal events. The leading irreducible backgrounds include 
\begin{itemize}
\item	 $\wbm Z(\gamma^{*})+$ QCD jets,
\item	 $\wbm \wbp \wbm +$ QCD jets, 
\end{itemize}
when the $W,Z$ bosons decay leptonically. We have ignored the faked leptons from heavy quarks like $b,c$ assuming that our stringent 
separation requirement for the charged leptons will effectively remove those. 
We have calculated the background processes using Madgraph \cite{Alwall:2007st}. In Table \ref{X2X2sigcut2}, we have listed the total cross sections for the signal as well as the leading backgrounds, after the basic cuts and after mass cuts for 7 and 14 TeV.
The reconstruction procedure outlined above effectively select out the signal kinematics, and substantially suppress the SM backgrounds.

\begin{table}[t]
	\centering
		\begin{tabular}{|c||c|c|c|} \hline
		\multicolumn{4}{|c|}{$\sigma (fb)$ } \\\hline
		    $\sqrt{s}$ &Process  &

		    Basic Cuts &

	        Cut on Masses
\\\hline
\multirow{3}{*}{ 7 TeV }&
	         $X^{--} X^{++}$  &

		    5.0&

			4.9
\\ &
		    $\lm\lp\wbm+$  2 QCD jets   &

			 $ 7.4$ &

         $1.4$
\\  &
		    $\wbm\wbp\wbm+$  2 QCD jets &

			 $ 0.022$ &
	         $0.0035 $\\\hline

\multirow{3}{*}{ 14 TeV }&
	         $X^{--} X^{++}$  &

		    13&

			13
\\ &
		    $\lm\lp\wbm+$  2 QCD jets   &

			 $ 30$ &

         $ 5.5 $
\\  &
		    $\wbm\wbp\wbm+$  2 QCD jets &

			 $ 0.12$ &
	         $0.018 $\\\hline

		\end{tabular}
	\caption{ \footnotesize Effects of the kinematical cuts on the production cross section at the LHC for the signal  $X^{--} X^{++} \to \lp \lm \lm\nub + $ 2 jets. $M=200$ GeV is assumed.}
	\label{X2X2sigcut2}
\end{table}

\newpage

\subsubsection{Significance versus Luminosity}

\begin{figure}[t]
	\includegraphics[scale=1, width=8cm]{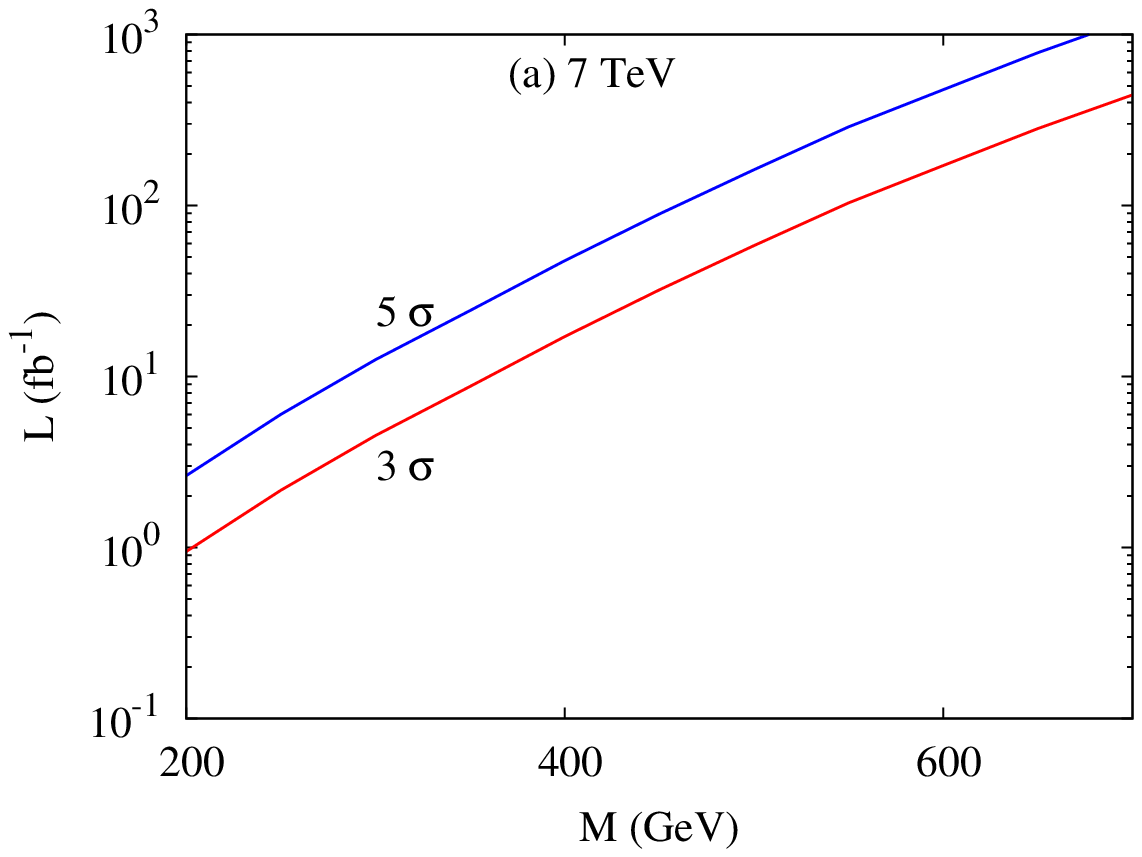}
	\includegraphics[scale=1, width=8cm]{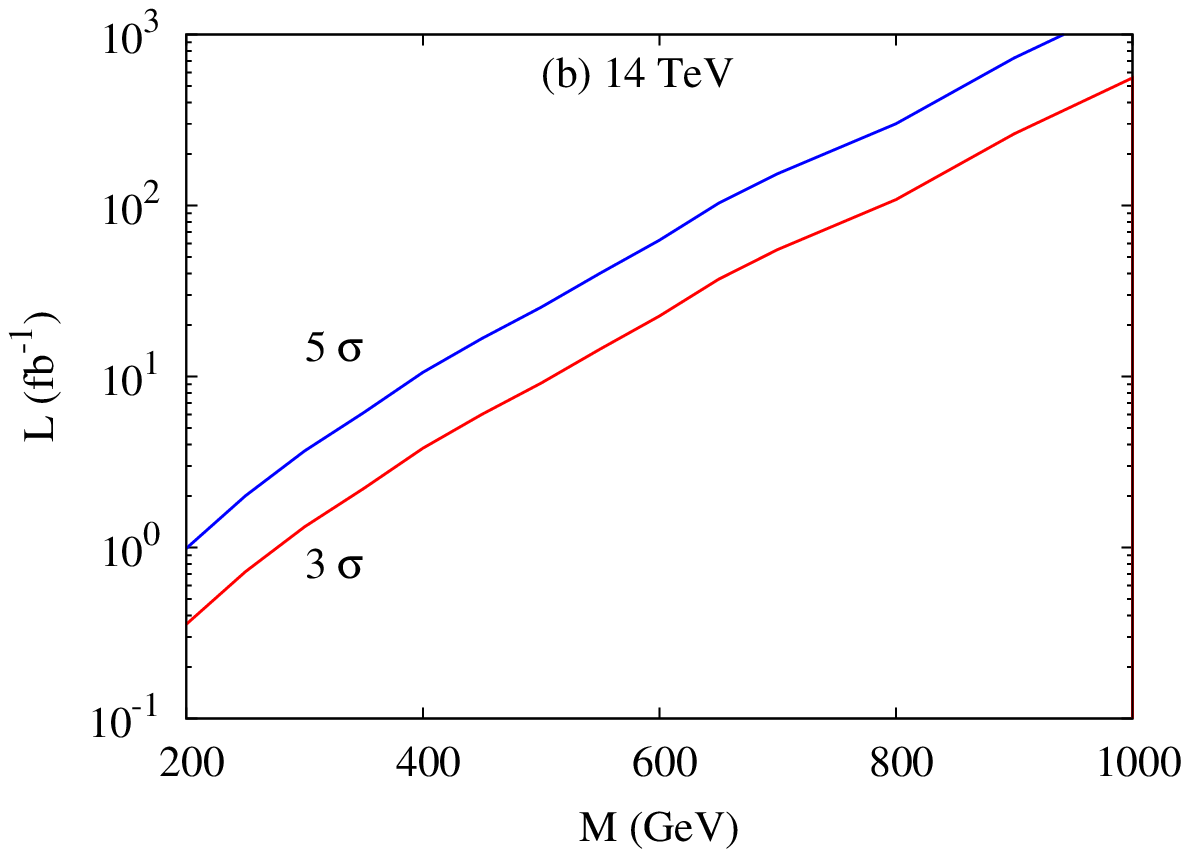}  
	\caption{Luminosity versus triplet mass for significances of 3$\sigma$ and 5$\sigma$ for the LHC (a) at 7 TeV and (b) 14 TeV.}	
     \label{luminosities}
\end{figure}

To quantify the signal observability, the statistical significance $s$ is conservatively defined as
\begin{equation}
	s = \frac{N_s}{\sqrt{N_s+N_b}} ,
\end{equation}
where $N$ refers to the number of events, and the subscripts $s$ and $b$ refers to the signal and the background respectively. If $L$ is the 
integrated luminosity and $\sigma$ the cross section, we can solve for the luminosity as a function of the significance
\begin{equation}
	L = s^2 \left( \frac{\sigma_s + \sigma_b}{\sigma_s^2} \right).
	\label{lumeq} 
\end{equation}
This allows us to calculate the luminosity needed to reach a given statistical significance. We extend the analysis in the last section 
including both $\ell^{-}\ell^{-}\ell^{+}$ and $\ell^{+}\ell^{+}\ell^{-}$ signal events, 
and use the events produced by Madgraph for the backgrounds discussed earlier.
After the kinematical and mass cuts are applied assuming a $60$ GeV mass window around the triplet mass. We present our results in Fig.~\ref{luminosities} for the 3$\sigma$ and 5$\sigma$ statistical significance at the energies of 7 TeV and 14 TeV.

\section{Conclusions}

The impressive experiments at the LHC have taken us to the energy and luminosity frontier for discovery of new particles. Among many exciting
new physics scenarios, the extension of the leptonic sector remains to be a well-motivated possibility due to the need for neutrino
mass. Similarly, models of gauge-Higgs unification also suggest the existence of exotic leptons. 

The most general phenomenological model involving a lepton triplet with hypercharge $\pm 1$ was constructed. A distinctive feature of this model is the prediction of a doubly charged lepton, and a heavy Dirac neutrino. We have carefully studied the coupling of these exotic leptons to gauge bosons, the Higgs and the SM leptons.
We then studied the phenomenology of these exotic leptons in low-energy experiments such as muon rare decays, tau rare decays or $Z$ boson FCNC decays. Using current experimental constraints, we obtained upper bounds on the mixing angles of the order of $10^{-3}$. We also consider constraints from the non-unitarity of the PMNS matrix, but we found that they are not as strong as the ones from FCNC decays. 
After this, we studied all the possible decay channels for the exotic leptons and the corresponding partial widths and branching fractions. 

We found that the exotic leptons can be pair-produced at the LHC with a cross section of 1 pb$-10^{-3}$ pb for a mass around 1 TeV. We propose to identify the doubly charged lepton via the channel $\ell^- \ell^- \nu \ell^+ j j$. After the selective acceptance cuts and kinematical reconstruction, we found that, 
if the mixing parameters between the exotic and light leptons are at the order of $10^{-6}$ or larger, then
their signal can be observable to a $3-5\sigma$ statistical significance for a 400 GeV mass at the 7 TeV LHC with $10-50$ fb$^{-1}$ luminosity, and for 800 GeV at the 14 TeV LHC with $100-300$ fb$^{-1}$ luminosity.

We would like to comment that the analysis done here is rather conservative and does not take into account many combinations of other channels
for production and decays. Therefore the prospects of discovery of these exotic particles may be improved with further analysis.

\subsection*{Acknowledgments}
We would like to thank Ian Lewis for discussions, and both referees for their valuable suggestions. We also acknowledge the Aspen Center for Physics for hospitality 
when a part of this work was carried out. 
The work of C.G.~and T.H. is supported in part by the U.S. Department of Energy under grant No. DE-FG02-95ER40896.
The work of A.D. is supported in part by the National Science Foundation under grant PHY-0905383-ARRA.
\appendix
\section{Change from weak to mass eigenstates}

The weak eigenstates and mass matrices can be written as
\begin{eqnarray}
	N = \begin{pmatrix} \nu_i \\ X^0 \end{pmatrix}, &\hfill& 
	E = \begin{pmatrix} e_i \\ X^-  \end{pmatrix}\\
	M_N =\begin{pmatrix} m_2 &\hfill& m_3^\dagger\\
	                             0    &\hfill&  M_1 \end{pmatrix}, &\hfill&
	M_E =\begin{pmatrix} m_1  &\hfill& \frac{m_3^\dagger}{\sqrt{2}} \\
	                             0    &\hfill&  M_1 \end{pmatrix}.	
	\label{eqA2}
\end{eqnarray}
Here, $m_1$ and $m_2$ are arbitrary 3x3 matrices, and $m_3$ is a row vector. 

The mass matrices (\ref{M}) can be diagonalized by biunitary transformations
\begin{eqnarray}
	S_E^\dagger M_E T_E &=& M_{Ed} = diag(m_e, m_\mu, m_\tau, M_2), \ \ S_N^\dagger M_N T_N = M_{Nd} = diag(m_{\nu_1}, m_{\nu_2}, m_{\nu_3}, M_3),~~~~
	\label{SDTap}
\end{eqnarray}
where, as usual, the matrices $S_E$, $T_E$, $S_N$ and $T_N$ are unitary and the diagonal elements are the tree level masses. 

Similar to a general fermionic sector with arbitrary Yukawa couplings, there are more theory parameters than those that can be
experimentally determined. Thus it is necessary to parameterize the fermionic sector by a few more physical parameters. 
Furthermore, we find convenient to introduce
\begin{eqnarray}
	v_E &=& S_E^\dagger \begin{pmatrix} 0 \\1 \end{pmatrix} \hspace{20pt} 	
	v_N = S_N^\dagger \begin{pmatrix} 0 \\1 \end{pmatrix} \label{vdef} \hspace{20pt}
	V = S_N^\dagger S_E.
\end{eqnarray}
Here, $0$ is a column vector with three vanishing components. Clearly, $V$ is a unitary matrix, and $v_E$ and $v_N$ are vectors of norm 1. Due to the particular form of Eq.~(\ref{eqA2}), these parameters are not independent. There are some relations among them, in fact $v_N = V v_E$. The first relation comes from the fact that
\begin{eqnarray}
	M_E^\dagger \begin{pmatrix} 0 \\ 1 \end{pmatrix} &=&  
	\begin{pmatrix} m_1^\dagger  &\hfill& 0 \\
    \frac{m_3}{\sqrt{2}}    &\hfill&  M_1 \end{pmatrix} \begin{pmatrix} 0 \\ 1 \end{pmatrix}= M_1 \begin{pmatrix} 0 \\ 1 \end{pmatrix}, \hspace{10pt}
	M_N^\dagger \begin{pmatrix} 0 \\ 1 \end{pmatrix} =  
	\begin{pmatrix} m_1^\dagger  &\hfill& 0 \\
    m_3    &\hfill&  M_1 \end{pmatrix} \begin{pmatrix} 0 \\ 1 \end{pmatrix}= M_1 \begin{pmatrix} 0 \\ 1 \end{pmatrix},
   \label{p1}
\end{eqnarray}
which due to Eqs.~(\ref{SDTap}) and (\ref{vdef}) implies
\begin{eqnarray}
	T_E^\dagger \begin{pmatrix} 0 \\ 1 \end{pmatrix} = \frac{1}{M_1} M_{Ed} v_E, \hspace{25pt}
	T_N^\dagger \begin{pmatrix} 0 \\ 1 \end{pmatrix} = \frac{1}{M_1} M_{Nd} v_N,
	\label{c1N}
\end{eqnarray}
which simplifies further to 
\begin{eqnarray}
	\vert v_{E4} \vert =M_1/M_2, \hspace{20pt}
	\vert v_{N4} \vert =M_1/M_3,
	\label{ue}
\end{eqnarray}
because the exotic leptons are very heavy compared to the leptons of the Standard Model.

The latter quantities correspond to the components of unitary vectors, therefore $M_1$ should be smaller than $M_2$ and $M_3$. In other words, the doubly charged lepton is lighter than the singly charged and neutral ones at \textit{tree} level. This fact is not true once quantum corrections are taken into account. 

The second set of relation comes from the fact that
\begin{equation}
	M_E \begin{pmatrix} 0 \\1 \end{pmatrix} =\begin{pmatrix} \rd m_3^\dagger \\M_1 \end{pmatrix}
		=  \begin{pmatrix} \rd & \hfill & 0 \\ 0  & \hfill & 1 \end{pmatrix} \begin{pmatrix} m_3^\dagger \\M_1 \end{pmatrix} = \begin{pmatrix} \rd & \hfill & 0 \\ 0  & \hfill & 1 \end{pmatrix}  M_N \begin{pmatrix} 0 \\1 \end{pmatrix}.
\end{equation}
Using Eqs.~(\ref{SDTap}), (\ref{c1N}) this can be rewritten as
\begin{equation}
	S_E M_{Ed}^2 v_E = \begin{pmatrix} \rd & \hfill & 0 \\ 0  & \hfill & 1 \end{pmatrix}  S_N M_{Nd}^2 v_N.
\end{equation}
Since 
\begin{eqnarray}
	 S_N^\dagger \begin{pmatrix} 0 & \hfill & 0 \\ 0  & \hfill & 1 \end{pmatrix}  S_N = v_N v_N^\dagger,
\end{eqnarray}
the relation can be written as 
\begin{equation}
	V M_{Ed}^2 v_E = \left(\rd + (1-\rd) v_N v_N^\dagger \right) M_{Nd}^2 v_N.
	\label{trivsdou}
\end{equation}
After breaking this equation into components, neglecting the mass of the SM leptons with respect to $M_1$, $M_2$ and $M_3$, and using the Eqs.~(\ref{ue}), this simplifies to
\begin{equation}
	V_{i4} = \left( \rd \frac{M_3^2}{M_1^2} \delta_{i4} + 1 - \rd \right) v^*_{E4} v_{Ni}.
\end{equation}
This equation implicitly expresses the vector $v_N$ in terms of $V_{i4}$. After multiplying this expression by $V^\dagger$ and using again Eqs.~(\ref{ue}) we get 
\begin{equation}
	V_{4i}^* = \sqrt{2} \left(\frac{M_2^2}{M_1^2} \delta_{i4} - 1+\rd \right) v_{N4}^* v_{Ei}.
\end{equation}
Similarly, this equation implicitly expresses the vector $v_E$ in terms of $V_{4i}$.

\section{Electroweak couplings}
The weak interactions are described by the lagrangian
\begin{eqnarray} \nonumber
	{\cal L}_{Weak} &=& \frac{g}{\cos \tw} Z^\mu J_\mu^N + \frac{g}{\sqrt{2}} (W^{+\mu} J_{\mu} +h.c), \\
	J_\mu^N &=& \overline{\Psi'_i} \gamma_\mu (T^3 - \sin^2 \tw Q ) \Psi'_i \hspace{20pt} J_\mu^+ = \overline{\Psi'_i} \gamma_\mu T^+ \Psi'_i,
\end{eqnarray}
where $i$ runs through all the leptonic particle content. The prime on the fields indicate that they are weak eigenstates. Since
\begin{eqnarray} \nonumber
	\mbox{For singlets} \hspace{10pt}&&T^+=0 \hspace{10pt} T^3 = 0\\
	 \nonumber
	 \mbox{For doublets} \hspace{10pt} &&
	 		T^+= \begin{pmatrix} 0 & \hfill & 1 \\ 0 & \hfill & 0 \end{pmatrix} \hspace{10pt} 
	 		T^3= \begin{pmatrix} \frac{1}{2} & \hfill & 0 \\ 0 & \hfill & -\frac{1}{2} \end{pmatrix}\\
	 \mbox{For Triplets }\hspace{10pt} && 
			T^+= \begin{pmatrix} 	0 & \hfill & \sqrt{2} & \hfill &1 \\ 
									0 & \hfill & 0 & \hfill & \sqrt{2} \\
					        		0 & \hfill & 0 & \hfill & 0  \\ \end{pmatrix} \hspace{10pt}
			T^3= \begin{pmatrix} 	1 & \hfill & 0 & \hfill & 0 \\ 
									0 & \hfill & 0 & \hfill & 0 \\
					        		0 & \hfill & 0 & \hfill & -1 \\ \end{pmatrix},
\end{eqnarray}
we get
\begin{eqnarray} 
	 J^N_\mu &=& \frac{1}{2} \overline{\nu'}_{Li} \gamma_\mu \nu'_{Li}+
				 (-\frac{1}{2} + \sin^2 \tw ) \overline{e'}_{Li} \gamma_\mu e'_{Li}+
				 \sin^2 \tw \overline{e'}_{Ri} \gamma_\mu e'_{Ri}+
			     \overline{X'^0_L} \gamma_\mu X'^0_L+
				 \overline{X'^0_R} \gamma_\mu X'^0_R  \nonumber\\
			   &&+ \sin^2 \tw \overline{X'^-_L} \gamma_\mu X'^-_L+
				   \sin^2 \tw \overline{X'^-_R} \gamma_\mu X'^-_R+
				 (-1+ 2 \sin^2 \tw ) \overline{X'^{--}_R} \gamma_\mu X'^{--}_R\nonumber\\
				 \nonumber
 			  &&+(-1+ 2 \sin^2 \tw ) \overline{X'^{--}_L} \gamma_\mu X'^{--}_L\\
	 J^+_\mu &=& \overline{\nu'}_{Li} \gamma_\mu e'_{Li} + 
	 			\sqrt{2} \hspace{2pt} \overline{X'^0_L} \gamma_\mu X'^-_L + 
	 			\sqrt{2} \hspace{2pt} \overline{X'^0_R} \gamma_\mu X'^-_R + 
	 			\sqrt{2} \hspace{2pt} \overline{X'^-_L} \gamma_\mu X'^{--_L} + 
	 			\sqrt{2} \hspace{2pt} \overline{X'^-_R} \gamma_\mu X'^{--_R}.
\end{eqnarray}
In terms of definitions (\ref{Col}), we have that
\begin{eqnarray} 
J^N_\mu &=& \overline{N'_L} \gamma_\mu \begin{pmatrix} \frac{1}{2} & \hfill & 0 \\ 0 & \hfill & 1\end{pmatrix}N'_L+
		    \overline{N'_R} \gamma_\mu \begin{pmatrix}           0 & \hfill & 0 \\ 0 & \hfill & 1\end{pmatrix}N'_R+
\nonumber	\overline{E'_L} \gamma_\mu  
	  		\begin{pmatrix} - \frac{1}{2} + \sin^2 \tw & \hfill & 0 \\ 0 & \hfill & \sin^2 \tw\end{pmatrix}E'_L\\ \nonumber
         &&+ \sin^2 \tw \overline{E'_R} \gamma_\mu E'_R +(-1+ 2 \sin^2 \tw) \overline{X'^--} \gamma_\mu X'^{--} \\
J^+_\mu &=& \overline{N'_L} \gamma_\mu \begin{pmatrix} 1 & \hfill & 0 \\ 0 & \hfill &  \sqrt{2} \end{pmatrix}E'_L+
\nonumber  	\overline{N'_R} \gamma_\mu \begin{pmatrix} 0 & \hfill & 0 \\ 0 & \hfill &  \sqrt{2} \end{pmatrix}E'_R\\
         &&+\overline{E'_L} \gamma_\mu \begin{pmatrix} 0  \\ \sqrt{2} \end{pmatrix} X'^{--}_L+
            \overline{E'_R} \gamma_\mu \begin{pmatrix} 0  \\ \sqrt{2} \end{pmatrix} X'^{--}_R.          
\end{eqnarray}
The first entry of the these matrices is a 3x3 matrix, corresponding to the three generations of the SM; and the last entry is a c-number. The task now is to express these currents as a function of the mass eigenstates. The mass eigenstates (the fields with no prime), according to Eq. ~(\ref{SDT}), are given by
\begin{eqnarray} 
	 E_L = S^\dagger_E E'_L \hspace{20pt} E_R = T^\dagger_E E'_R \hspace{20pt} N_R =T^\dagger_E N'_R \nonumber \hspace{20pt} N_L = S^\dagger_N N'_L. \nonumber
\label{WeToMa}
\end{eqnarray}
Using this, and the fact that $\Psi_L = \frac{1}{2} (1-\gamma_5) \Psi$ and $\Psi_R = \frac{1}{2} (1+\gamma_5) \Psi$, the weak currents can be expressed as
\begin{eqnarray}
	J^N_\mu &=& \overline{N} \gamma_\mu (A - B \gamma_5) N +
				\overline{E} \gamma_\mu (C - D \gamma_5) E -
				(-1+ 2 \sin^2 \tw) \overline{X'^--} \overline{X^{--}} \gamma_\mu X^{--}\\
	J^+_\mu &=& \overline{N} \gamma_\mu (F - G \gamma_5) E +
				\overline{E} \gamma_\mu (H - J \gamma_5) X^{--},
			\label{defABC}
\end{eqnarray}
where
\begin{eqnarray}
	A &=& S^\dagger_N diag \left(\frac{1}{4}, \frac{1}{2} \right) S_N + 
		  T^\dagger_N diag \left(0, \frac{1}{2} \right) T_N \hspace{5pt},		
	B = S^\dagger_N diag \left(\frac{1}{4}, \frac{1}{2}\right) S_N - 
		  T^\dagger_N diag \left(0, \frac{1}{2} \right) T_N \nonumber\\
	C &=& S^\dagger_E diag \left(-\frac{1}{4}+ \frac{1}{2} \sin^2 \tw , \frac{1}{2} \sin^2 \tw\right) S_E  
		  + \frac{1}{2} \sin^2 \tw \nonumber\\		
	D &=& S^\dagger_E diag \left(-\frac{1}{4}+ \frac{1}{2} \sin^2 \tw , \frac{1}{2} \sin^2 \tw\right) S_E  
		  - \frac{1}{2} \sin^2 \tw\\	
	F &=& S^\dagger_N diag \left(\frac{1}{2}, \rd\right) S_E + 
		  T^\dagger_N diag \left(0, \rd \right) T_E, \hspace{5pt}		
  	G = S^\dagger_N diag \left(\frac{1}{2}, \rd \right) S_E - 
  		  T^\dagger_N diag \left(0, \rd \right) T_E\nonumber\\
  	H &=& \frac{1}{2} (S_E+T_E)^\dagger \begin{pmatrix} 0\\ \sqrt{2} \end{pmatrix}, \hspace{10pt}					
  	J = \frac{1}{2} (S_E-T_E)^\dagger \begin{pmatrix} 0\\ \sqrt{2} \end{pmatrix}.\nonumber
\end{eqnarray}
These matrices can be expressed in terms of the unitary matrix $V$ and the vectors $v_E$ and $v_N$ (which can also be expressed in terms of V).  Using Eqs. (\ref{vdefap}) and (\ref{c1N}) , it is easy to show that
\begin{eqnarray}
	S^\dagger_N diag \left(a, b \right) S_N &=& a + (b-a) v_N v^\dagger_N \hspace{20pt} 
	S^\dagger_E diag \left(a, b \right) S_E = a + (b-a) v_E v^\dagger_E \nonumber\\
	T^\dagger_N diag \left(0, 1 \right) T_N &=& \frac{1}{M^2} M_{Nd}v_N v^\dagger_N M_{Nd} \hspace{20pt}
	T^\dagger_E diag \left(0, 1 \right) T_E = \frac{1}{M^2} M_{Ed}v_E v^\dagger_E M_{Ed}.	
\end{eqnarray}
As a result, the coupling matrices are
\begin{eqnarray}
	A &=& \frac{1}{4} \left(1 + v_N v^\dagger_N \right) + \frac{1}{M^2} M_{Nd}v_N v^\dagger_N M_{Nd} \hspace{15pt}
	B = \frac{1}{4} \left(1 + v_N v^\dagger_N \right) - \frac{1}{M^2} M_{Nd}v_N v^\dagger_N M_{Nd} \nonumber\\
	C &=& - \frac{1}{4} + \sin^2 \tw + \frac{1}{4} v_E v^\dagger_E \hspace{15pt}		
	D = - \frac{1}{4} +\frac{1}{4} v_E v^\dagger_E \nonumber \\	
	F &=& \frac{1}{2} \left(V + (\sqrt{2}-1) v_N v^\dagger_E \right) + \frac{1}{\sqrt{2}M^2} M_{Nd}v_N v^\dagger_E M_{Ed} \\
	G &=& \frac{1}{2} \left(V + (\sqrt{2}-1) v_N v^\dagger_E \right) - \frac{1}{\sqrt{2}M^2} M_{Nd}v_N v^\dagger_E M_{Ed} \nonumber\\
	H &=& \rd \left(v_E + \frac{1}{M} M_{Ed} v_E\right) \hspace{15pt}
	J = \rd \left(v_E - \frac{1}{M} M_{Ed} v_E\right). \nonumber
\end{eqnarray}
To get the actual couplings of the gauge fields to the leptons it is necessary to break the latter matrices in components and if necessary apply equations~(\ref{newcon2}) in order to simplify. The results are in the Table \ref{T1}.

The couplings that describe the interaction among $W^-$, $X^0$ and the SM model charged leptons are very interesting. These are $F_{4\alpha}$ and $G_{4\alpha}$, where $\alpha = e, \mu,\tau$. Notice that equations~(\ref{newcon2}) imply that $V_{4\alpha} = (-\sqrt2+1) v_{N4}v^*_{\alpha}$. As a result we have that:
\begin{eqnarray}
	F_{4\alpha} = -G_{4\alpha} = \frac{1}{\sqrt{2}M^2} ( M_{Nd}v_N v^\dagger_E M_{Ed} )_{4\alpha} = \frac{1}{\sqrt2} \frac{m_\alpha}{M} v_{N4} v_{\alpha}^* 
\end{eqnarray}
Hence they are highly suppressed. As a result the interaction among $W^-$, $X^0$ and the SM model charged leptons is neglectable. 
\newpage

\section{Generalization to an arbitrary gauge}

In an arbitrary gauge, Yukawa's interactions are described by the lagrangian
\begin{eqnarray}
-{\cal L}_{Y}={\lambda}_1\overline{L'_L}He'_R+{\lambda}_2\overline{L'_L}H^c{\nu'}_R+{\lambda}_3\overline{X'}_RH^cL'_L+M_1\overline{X'}X'+h.c.
\end{eqnarray}
The prime on the fields indicate they are weak eigenstates. We also follow the notation
 \begin{eqnarray}
 H=\left(\begin{array}{c}
 {\phi}^+\\
 \frac{1}{\sqrt{2}}(v+h(x)+i\eta(x))
 \end{array}\right),
 H^c=\left(\begin{array}{c}
  \frac{1}{\sqrt{2}}(v+h(x)-i\eta(x))\\
 -{\phi}^{+*}

 \end{array}\right)
 \end{eqnarray}

\begin{eqnarray}
 N=\left(\begin{array}{c}
 \nu\\
 X^0
 \end{array}\right),
 E=\left(\begin{array}{c}
  e\\
  X^-
 \end{array}\right),
 M_N=\left(\begin{array}{cc}
 m_2&m_3^*\\
 0& M_1
 \end{array}\right),
 M_E=\left(\begin{array}{cc}
  m_1&\frac{m_3^*}{\sqrt2}\\
 0& M_1
 \end{array}\right).
 \end{eqnarray}
Setting $m_i=\frac{\lambda_i v}{\sqrt2}$, we can rewrite the lagrangian as
\begin{eqnarray}
 -{\cal L}_ {Y}&=&\overline{E'_L} M_EE'_R+\overline{N'_L} M_NN'_R+\frac{h(x)}{v}\overline{E'_L} M_E|_{M_1=0}E'_R
 +\frac{h(x)}{v}\overline{N'_L} M_N|_{M_1=0}N'_R\nonumber\\
 &+&i\frac{\eta(x)}{v}\overline{E'_L} M_E|_{M_1=0}E'_R-i\frac{\eta(x)}{v}\overline{N'_L} M_N|_{M_1=0}\left(\begin{array}{cc}
 1&0\\
 0& -1
 \end{array} \right)N'_R
 +\frac{{\phi}^+}{u}\overline{N'_L} M_E|_{M_1=0}\left(\begin{array}{cc}
 1&0\\
 0& -1
 \end{array}\right)E'_R\nonumber\\
&-&\sqrt2\frac{{\phi}^{+*}}{v}\overline{E'_L} M_N\left(\begin{array}{cc}
 1&0\\
 0& 0
 \end{array}\right)N'_R
 -2\frac{{\phi}^+}{v}\overline{E'_L} M_E|_{M_1=0}\left(\begin{array}{c}
 0\\
 1
 \end{array}\right)X'^{--}_R+h.c.
\end{eqnarray}
Using Eq. (\ref{SDTap}), (\ref{vdef}) and (\ref{WeToMa}), we can write this lagrangian in terms of the mass eigenstates
\begin{eqnarray}
 -{\cal L}_{Y}&=&\overline{E_L} M_{Ed}E_R+\overline{N_L} M_{Nd}N_R+\frac{h(x)}{v}\overline{E_L}\left(1-v_Ev_E^{\dagger}\right) M_{Ed}E_R
 +\frac{h(x)}{v}\overline{N_L}\left(1-v_Nv_N^{\dagger}\right) M_{Nd}N_R\nonumber\\
 &+&i\frac{\eta(x)}{v}\overline{E_L}\left(1-v_Ev_E^{\dagger}\right) M_{Ed}E_R
-i\frac{\eta(x)}{v}\overline{N_L}\left(1- v_N v_N^{\dagger}\right) M_{Nd} \left(1- 2 \frac{M^2_{Nd}}{M^2} v_N v_N^{\dagger} M_{Nd}\right) N_R \nonumber\\
&+&\sqrt2\frac{{\phi}^+}{v}\overline{N_L} V \left(1- v_E v_E^\dagger\right) M_{Ed} \left(1- 2\frac{M_{Ed}} v_Ev_E^{\dagger} M_{Ed}\right) E_R \nonumber\\
&-& \sqrt2 \frac{{\phi}^{+*}}{v} \overline{E_L} V^{\dagger} \left(1- v_N v_N^\dagger\right) M_{Nd}\left(1- \frac{M_{Nd} v_N v_N^\dagger M_{Nd}}{M^2}\right) N_R \nonumber\\
&-& 2\frac{{\phi}^+}{v}\overline{E_L}\left(\frac{M^2_{Nd}}{M}-M\right)v_EX^{--}_R +h.c. 
\end{eqnarray}
Clearly, the couplings of the Goldstone bosons to the standard model leptons are negligible compared to the couplings to the exotic leptons. In fact, using relations (\ref{newcon}) and the fact that $M_W = \frac{gv}{2}$, it is easy to show that the interaction lagrangian of the Goldstone bosons is
\begin{eqnarray}
{\cal L} &=& \dfrac{gM}{2M_W} \sum_i \left( (h+i\eta)(v_{Ni} \overline{\nu_{iL}} X^0_R + v_{Ei} \overline{e_{iL}} X^-_R ) - \phi^+ (v_{Ni} \overline{\nu_{iL}}X^-_R + 2 v_{Ei} \overline{e_{iL}} X^{--}_R )+ h.c.\right).~~~~
\end{eqnarray}
It is remarkable that the coupling of the Goldstone boson $\phi^+$ to exotic leptons is proportional to their charge. In particular, the coupling of $\phi^+$ to $X^0$ vanishes. The $W^+$ boson, as table \ref{T1} shows, does not couple to $X^0$ either, in agreement with equivalence theorem.  

\newpage

\section{$l_2 \to l_1 \gamma$ decays}

\subsection{$\mu \to e \gamma$ decay}

\begin{picture}(300,340)(-50,0)
\ArrowLine(30,270)(60,270)\ArrowLine(60,270)(120,270)\ArrowLine(120,270)(150,270)
\PhotonArc(90,270)(30,0,180){1}{18} \Photon(90,300)(90,330){1}{6}
\Text(30,277)[]{${\mu}^-$}\Text(95,277)[]{$\nu,X^0$}
\Text(155,277)[]{$e^-$} \Text(60,292)[]{$W^-$}
\Text(125,292)[]{$W^-$}\Text(94,330)[]{$\gamma$}
\Text(91,260)[]{(1)}
\ArrowLine(180,270)(210,270)\ArrowLine(210,270)(270,270)\ArrowLine(270,270)(300,270)
\PhotonArc(240,270)(30,90,180){1}{9}\DashCArc(240,270)(30,0,90){3}
\Photon(240,300)(240,330){1}{6}
\Text(180,277)[]{${\mu}^-$}\Text(240,277)[]{$\nu,X^0$}
\Text(305,277)[]{$e^-$} \Text(210,292)[]{$W^-$}
\Text(275,292)[]{${\phi}^-$}\Text(245,330)[]{$\gamma$} 
\Text(241,260)[]{(2)}
\ArrowLine(30,180)(60,180)\ArrowLine(60,180)(120,180)\ArrowLine(120,180)(150,180)
\DashCArc(90,180)(30,90,180){3}\PhotonArc(90,180)(30,0,90){1}{9}\Photon(90,210)(90,240){1}{6}
\Text(30,187)[]{${\mu}^-$}\Text(95,187)[]{$\nu,X^0$}
\Text(155,187)[]{$e^-$} \Text(60,202)[]{${\phi}^-$}
\Text(125,202)[]{$W^-$}\Text(94,240)[]{$\gamma$}
\Text(91,170)[]{(3)}
\ArrowLine(180,180)(210,180)\ArrowLine(210,180)(270,180)\ArrowLine(270,180)(300,180)
\DashCArc(240,180)(30,0,180){3} \Photon(240,210)(240,240){1}{6}
\Text(180,187)[]{${\mu}^-$}\Text(240,187)[]{$\nu,X^0$}
\Text(305,187)[]{$e^-$} \Text(210,202)[]{${\phi}^-$}
\Text(275,202)[]{${\phi}^-$}\Text(245,240)[]{$\gamma$}
\Text(241,170)[]{(4)}
\ArrowLine(30,90)(60,90)\ArrowLine(60,90)(120,90)\ArrowLine(120,90)(150,90)
\PhotonArc(90,90)(30,0,180){1}{18} \Photon(90,120)(90,150){1}{6}
\Text(30,97)[]{${\mu}^-$}\Text(95,97)[]{$X^{--}$}
\Text(155,97)[]{$e^-$} \Text(60,112)[]{$W^+$}
\Text(125,112)[]{$W^+$}\Text(94,150)[]{$\gamma$} 
\Text(91,80)[]{(5)}
\ArrowLine(180,90)(210,90)\ArrowLine(210,90)(270,90)\ArrowLine(270,90)(300,90)
\PhotonArc(240,90)(30,90,180){1}{9}\DashCArc(240,90)(30,0,90){3}
\Photon(240,120)(240,150){1}{6}
\Text(180,97)[]{${\mu}^-$}\Text(240,97)[]{$X^{--}$}
\Text(305,97)[]{$e^-$} \Text(210,112)[]{$W^+$}
\Text(275,112)[]{$\phi^+$}\Text(245,150)[]{$\gamma$}
\Text(241,80)[]{(6)}
\ArrowLine(30,0)(60,0)\ArrowLine(60,0)(120,0)\ArrowLine(120,0)(150,0)
\DashCArc(90,0)(30,90,180){3}\PhotonArc(90,0)(30,0,90){1}{9}
\Photon(90,30)(90,60){1}{6}
\Text(30,7)[]{${\mu}^-$}\Text(90,7)[]{$X^{--}$}
\Text(155,7)[]{$e^-$} \Text(60,22)[]{$\phi^+$}
\Text(125,22)[]{$W^+$}\Text(95,60)[]{$\gamma$}
\Text(91,-10)[]{(7)}
\ArrowLine(180,0)(210,0)\ArrowLine(210,0)(270,0)\ArrowLine(270,0)(300,0)
\DashCArc(240,0)(30,00,180){3}
\Photon(240,30)(240,60){1}{6}
\Text(180,7)[]{${\mu}^-$}\Text(240,7)[]{$X^{--}$}
\Text(305,7)[]{$e^-$} \Text(210,22)[]{$\phi^+$}
\Text(275,22)[]{$\phi^+$}\Text(245,60)[]{$\gamma$}
\Text(241,-10)[]{(8)}
\ArrowLine(30,-90)(60,-90)\ArrowLine(60,-90)(120,-90)\ArrowLine(120,-90)(150,-90)
\PhotonArc(90,-90)(30,0,180){1}{18}\Photon(90,-90)(90,-120){1}{6}
\Text(30,-83)[]{${\mu}^-$}\Text(95,-83)[]{$e,X^-$}
\Text(155,-83)[]{$e^-$}
\Text(95,-120)[]{$\gamma$},\Text(90,-50)[]{$Z^0$}
\Text(81,-132)[]{(9)}
\ArrowLine(180,-90)(210,-90)\ArrowLine(210,-90)(270,-90)\ArrowLine(270,-90)(300,-90)
\DashCArc(240,-90)(30,0,180){3}\Photon(240,-90)(240,-120){1}{6}
\Text(180,-83)[]{${\mu}^-$}\Text(245,-83)[]{$e,X^-$}
\Text(305,-83)[]{$e^-$}
\Text(245,-120)[]{$\gamma$},\Text(235,-50)[]{$h,\eta$}
\Text(231,-132)[]{(10)}
\ArrowLine(30,-180)(60,-180)\ArrowLine(60,-180)(120,-180)\ArrowLine(120,-180)(150,-180)
\PhotonArc(90,-180)(30,0,180){1}{18}\Photon(90,-180)(90,-210){1}{6}
\Text(30,-173)[]{${\mu}^-$}\Text(90,-173)[]{$X^{--}$}
\Text(155,-173)[]{$e^-$}
\Text(95,-210)[]{$\gamma$},\Text(90,-142)[]{$W^+$}
\Text(81,-220)[]{(11)}
\ArrowLine(180,-180)(210,-180)\ArrowLine(210,-180)(270,-180)\ArrowLine(270,-180)(300,-180)
\DashCArc(240,-180)(30,0,180){3}\Photon(240,-180)(240,-210){1}{6}
\Text(180,-173)[]{${\mu}^-$}\Text(240,-173)[]{$X^{--}$}
\Text(305,-173)[]{$e^-$}
\Text(245,-210)[]{$\gamma$},\Text(240,-142)[]{${\phi}^+$}
\Text(231,-220)[]{(12)}
\end{picture}

\vspace{240pt}
Using Lorentz invariance and Gordon decomposition, it is possible to show that the amplitude can always be written as
\begin{eqnarray}
 T(\mu \to e \gamma)=A \overline{u}_e(p-q)(1+{\gamma}_5)(2 p\cdot \epsilon-m_{\mu}{\not\!\epsilon})u_{\mu}(p),
\end{eqnarray}
where $\epsilon$ is the polarization of photon, $q$ is its momentum, $p$ is the momentum of the muon and $A$ is a constant. We can rewrite $A$ in terms of a dimensionless quantity $\delta$
\begin{eqnarray}
	A = \frac{eg^2 m_\mu}{256 M_W^2 \pi^2 } \delta.
\end{eqnarray}
It can be shown ~\cite{chengli} that the branching fraction is given then by
\begin{eqnarray}
	Br ( \mu \to e \gamma) = \frac{3 \alpha}{32 \pi} | \delta |^2.
\end{eqnarray}
In order to obtain $A$, our strategy will be to isolate the $p\cdot \epsilon$ term in our calculation. For simplicity, we work in the Feynman gauge. We also find convenient to introduce the notation $ r_a = \frac{M^2_a}{M^2}$.

\noindent
{\it Diagrams 1, 2, 3 and 4:} If they have only SM particles and no neutral flavor changing vertices, they are proportional to $(\frac{m_{\nu_i}}{m_W})^2 \sim 10^{-22}$ (due to GIM mechanism ~\cite{Abada:2008ea,chengli}). Furthermore, as we will see, all the other contributions are proportional to $v_e^* v_\mu$ to leading order . Since we hope to see the exotic leptons at the LHC, we neglect these contributions, otherwise we would have to assume that $v_e^* v_\mu$ is very small. 

Similarly,  since the coupling among the light charged leptons, the neutral exotic lepton and $W^+$ or $\phi^+$ is highly suppressed (as shown in table \ref{T1} and in appendix C), diagrams 1-4 with exotic leptons are proportional to $m_\mu m_e/M^2$. As a result, for our purposes we can assume
\begin{eqnarray}
	\sum^4_{i=1} A_i = 0.\nonumber
\end{eqnarray}

\noindent
{\it Diagrams 5, 6, 7 and 8:} A careful inspection of the diagrams shows that the only difference among these and diagrams 1,2 3 and 4 are the coupling of the leptons to $\phi^+$ or $W^+$ and the electric charge sign of the boson on the loop. Diagrams 1,2,3 and 4 would be the contribution of a exotic neutrino to the process $ \mu \to e \gamma$. This has been studied many times (see ~\cite{Abada:2008ea,chengli}). We follow the notation of \cite{chengli}, thus according to the results on table \ref{T1} and appendix C, we just have to take the coupling $U_{i}$ of the exotic neutrino to the leptons as $U_{i} \to \sqrt{2}v_{Ei}$ and flip the sign of the electric charge. After doing this, we get
\begin{eqnarray}
	\sum^8_{i=5} A_i= + \dfrac{e g^2 m_\mu }{32 \pi^2 M_W^2} g (r_W) v^*_{e} v_\mu,\nonumber
\end{eqnarray}
where
\begin{eqnarray}
	g(x) & = & \int^1_0 \dfrac{1-\alpha}{(1-\alpha) x + \alpha } \left[ 2(1-\alpha)(2-\alpha) x + \alpha (1+ \alpha)  \right] d\alpha \nonumber\\
	     & = & \frac{2}{3} - \frac{3x^3}{(1-x)^3} -\frac{15}{2} \left(\frac{x}{1-x}\right)^2 - \frac{11}{2}\left( \frac{x}{1-x}\right) + \frac{3 x \log x}{(1-x)^4}.
\end{eqnarray}

\noindent
{\it Diagrams 9, 10, 11 and 12:} Similarly, after a careful calculation we have found that
\begin{eqnarray}
	\sum^{12}_{i=9} A_i = \dfrac{e g^2 m_\mu }{256 \pi^2 M_W^2} \left( -8 + 12 \sin^2 \theta_W + 2 f(r_Z)+ \frac{1}{r_H} f(r_H) + \frac{1}{r_Z} f(r_Z) + 16 f(r_W) + \frac{8}{r_W} f(r_W) \right) v^*_{e} v_\mu ,\nonumber
\end{eqnarray}
where 
\begin{eqnarray}
	f(x) & = & \int^1_0 \frac{(1-\alpha) x}{ (1-\alpha) x+ \alpha } d\alpha  = -\frac{x}{1-x}\left( 1+ \frac{\log x}{1-x} \right).
\end{eqnarray}
The $M$-independent part corresponds to the contribution of the diagrams with no exotic leptons to leading order. 
Our final result is
\begin{eqnarray}
\delta = \left[-8 + 12 \sin^2 \theta_W + 8 g(r_W)+ \left(2 +\frac{1}{r_Z} \right) f(r_Z) + 8 \left(2 +\frac{1}{r_W} \right) f(r_W) + \frac{1}{r_H}  f(r_H) \right] v^*_{e} v_\mu. 
\end{eqnarray}

\subsection{$\tau \to l \gamma$ decay}

For this case it is a good approximation to assume that the final lepton is massless compared to $\tau$. We can also use the results of the previous section, with a slight modification to account for the additional hadronic channels of the $\tau$ decay. Hence

\begin{eqnarray}
\delta = \left[-8 + 12 \sin^2 \theta_W + 8 g(r_W)+ \left(2 +\frac{1}{r_Z} \right) f(r_Z) + 8 \left(2 +\frac{1}{r_W} \right) f(r_W) + \frac{1}{r_H}  f(r_H) \right] v^*_{l} v_\tau,
\end{eqnarray}
and
\begin{eqnarray}
	Br ( \tau \to l \gamma) = \left(\frac{G_F^2m_\tau^5}{192\pi^3\Gamma_\tau} \right) \left( \frac{3 \alpha}{32 \pi} | \delta |^2 \right),
\end{eqnarray}
where $\Gamma_\tau = 2.27 \times 10^{-12}$ GeV \cite{datagroup}.



\end{document}